\documentclass[iop]{emulateapj}
\newcommand{\oiii}{[O~{\footnotesize III}]}
\newcommand{\hii}{H~{\footnotesize II}}
\newcommand{\nii}{[N~{\footnotesize II}]}

\newcommand{\actaa}{Acta Astron.}
\newcommand{\nar}{New Astron. Rev.}
\def\etal{{et~al.\null}}
\shorttitle{Massive PN Central Stars from an Old Population}
\shortauthors{Davis et al.}

\begin{document}

\title{The True Luminosities of Planetary Nebulae in M31's Bulge: Massive Central Stars from an Old Stellar Population}
\author{Brian D. Davis, Robin Ciardullo\altaffilmark{1}}
\affil{Department of Astronomy \& Astrophysics, The Pennsylvania
State University, University Park, PA 16802}
\email{bdavis@psu.edu, rbc@astro.psu.edu}
\author{George H. Jacoby}
\affil{Lowell Observatory, Flagstaff, AZ 86001}
\email{gjacoby@lowell.edu}
\author{John. J. Feldmeier}
\affil{Department of Physics and Astronomy, Youngstown State University, Youngstown, OH 44555}
\email{jjfeldmeier@ysu.edu}
\and
\author{Briana L. Indahl}
\affil{Department of Astronomy, University of Texas, Austin, TX 78712}
\email{bindahl@utexas.edu}

\altaffiltext{1}{Institute for Gravitation and the Cosmos, The Pennsylvania State University, University Park, PA 16802}

\begin{abstract}
We measure the Balmer decrements of 23 of the brightest planetary nebulae (PNe) in the inner bulge ($r \lesssim 3\arcmin$) of M31 and de-redden the bright end of the region's  \oiii\ $\lambda 5007$ planetary nebula luminosity function. We show that the most luminous PNe produce $\gtrsim 1{,}200 \, \rm{L}_{\odot}$ of power in their \oiii\ $\lambda 5007$ line, implying central star luminosities of at least $\sim 11{,}000 \, \rm{L}_{\odot}$. Even with the most recent accelerated-evolution post-AGB models, such luminosities require central star masses in excess of $0.66 \, \rm{M}_{\odot}$, and main sequence progenitors of at least $\sim 2.5 \, \rm{M}_{\odot}$.   Since M31's bulge has very few intermediate-age stars, we conclude that conventional single-star evolution cannot be responsible for these extremely luminous objects.  We also present the circumstellar extinctions for the region's bright PNe and demonstrate that the distribution is similar to that found for PNe in the Large Magellanic Cloud, with a median value of $A_{5007} = 0.71$. Finally, we compare our results to extinction measurements made for PNe in the E6 elliptical NGC~4697 and the interacting lenticular NGC~5128. We show that such extinctions are not unusual, and that the existence of very high-mass PN central stars is a general feature of old stellar populations. Our results suggest that single-star population synthesis models significantly underestimate the maximum luminosities and total integrated light of AGB stars.
\end{abstract}

\keywords{planetary nebulae: general --- galaxies: individual (M31) --- galaxies: bulges --- galaxies: stellar content}

\section{Introduction}

Since the 1980's, the \oiii\ $\lambda 5007$ planetary nebula luminosity function (PNLF) has served as a reliable standard candle for determining galactic distances out to $\sim 20$ Mpc (beyond which the required exposure times become unreasonable, even for large ground-based telescopes).  \cite{jacoby1980} found that the faint end of the \oiii\ $\lambda 5007$ PNLF could be well-modeled by the exponential function expected from slowly evolving central stars embedded in rapidly expanding, optically thin nebulae \citep{henize1963}. Though we seem to understand some features of the PNLF, the detailed nature of its bright-end cutoff remains a mystery. \cite{ciardullo1989a} first suggested a cutoff exponential to model the bright end of the PNLF in Local Group galaxies:
\begin{equation}
N(m) \propto e^{0.307 m}\left[1-e^{3(m^* - m)}\right],
\label{eq:N(m)}
\end{equation}
where $N$ is the number of planetary nebulae (PNe), $m^*$ is the \oiii\ $\lambda 5007$ apparent magnitude that defines the function's bright-end cutoff, and $m$, the observed \oiii\ $\lambda 5007$ magnitude of each PN, is related to monochromatic flux (in erg~s$^{-1}$~cm$^{-2}$) by
\begin{equation}
m_{5007} = -2.5\log F_{5007} - 13.74.
\label{eq:m=F}
\end{equation}

\cite{ciardullo1989a} first estimated $M^*$, the \oiii\ $\lambda 5007$ absolute magnitude of the brightest possible PN, by using 104 objects in M31 and adopting the Cepheid distance to the galaxy.  Since then, the foreground extinction in the direction of M31 has been more accurately measured, and the results from other galaxies have been included.  The current estimated value of the PNLF cutoff is $M^* = -4.54 \pm 0.05$ \citep{ciardullo2013}.

Recent surveys have shown that the faint end of the PNLF can vary considerably, depending on the details of the stellar population \citep[e.g.,][]{jacoby2002, longobardi2015, hartke2018}. However, aside from the well-known fading of $M^*$ in low-metallicity populations \citep[e.g.,][]{ciardullo1992, ciardullo2002, hernandez2009, schonberner2010}, the PNLF cutoff has proven to be remarkably consistent across a very wide range of galaxy types and stellar populations, including spiral disks and bulges \citep[e.g.,][]{feldmeier1997, ciardullo2002}, elliptical and cD galaxies \citep[e.g.,][]{ciardullo1989b, jacoby1990}, irregular and interacting systems \citep[e.g.,][]{hui1993, mcmillan1993, johnson2009}, and even intracluster space \citep{longobardi2015}. A summary of the history of the PNLF is given in \cite{ciardullo2013}, and a more comprehensive review can be found in \cite{jacoby1992}.

One major drawback of the PNLF  is that, although the method has proven reliable, there is no theoretical framework which explains why it should work in such a diverse set of populations. In fact, there are good reasons to believe that the method should not work. As shown by a number of authors, and most recently by \cite{ciardullo2013}, the \oiii\ $\lambda 5007$ magnitude of the PNLF cutoff should be very sensitive to population age, so one would not expect $M^*$ to be a reliable standard candle. The argument is as follows:

As a stellar population ages, its main sequence turnoff mass decreases, and, through the initial-final mass relation \citep[e.g.,][]{kalirai2008, casewell2015}, so do the post-asymptotic giant branch (post-AGB) central stars that it produces.  Since the luminosities of these central stars depend very sensitively on their mass \citep{paczynski1970,vassiliadis1994}, the luminosities of their surrounding nebulae should fade dramatically with age \citep[e.g.,][]{marigo2004}. Based on the multitude of tests performed on the method, this fading clearly does not occur.

A related, and even more puzzling issue comes from the observed absolute magnitude of $M^*$.  A value of $M^* \approx -4.5$ implies that PNe near the PNLF cutoff emit $\sim 600 \, \rm{L}_{\odot}$ in their \oiii\ $\lambda 5007$ emission line.   Since, at most, $\sim 11\%$ of the emission of a PN's central star can be reprocessed into \oiii\ $\lambda 5007$ radiation \citep{dopita1992, schonberner2010, gesicki2018}, this means that the central stars of $M^*$ PNe must have luminosities in excess of $\sim 5{,}500 \, \rm{L}_{\odot}$.   According to models of post-AGB evolution, such luminosities can only be produced by high-mass ($\gtrsim 0.60 \, \rm{M}_{\odot}$) central stars  \citep{vassiliadis1994, blocker1995}, and this implies relatively high-mass ($\gtrsim 1.9 \, \rm{M}_{\odot}$) progenitors \citep[e.g.,][]{kalirai2008, casewell2015}.  These early F stars, which have main sequence lifetimes of only $\sim 1.5$~Gyr, should not be present in old stellar populations \citep{trager2000,jeong2012}.
	
Recently, \cite{bertolami2016} published a new set of accelerated-evolution post-AGB models that allow $0.57 \, \rm{M}_{\odot}$ PN central stars to achieve luminosities of $\sim 5{,}500 \, \rm{L}_{\odot}$, i.e., the same as those reached by $0.60 \, \rm{M}_{\odot}$ central stars in previous post-AGB models. Moreover, the stellar luminosities derived from these new models have a weaker dependence on central star mass than previous calculations, suggesting  that $M^*$ may not depend as much on population age.   This is a major step forward toward understanding the PNLF\null. Unfortunately, these models, like the previous generation of calculations, have one important limitation.
	
As AGB stars evolve, they lose mass from their outer envelopes at rates as high as $10^{-4} \, \rm{M}_{\odot}$~yr$^{-1}$
\citep[e.g.,][]{knapp1985, winters2000}. The ejected gas and dust enshroud the core, until eventually the star becomes hot enough to ionize much of this surrounding material, creating a PN\null. Unless all of the dust ejected from the AGB progenitor is radiatively destroyed, this material will serve to attenuate some fraction of the PN's radiation, thereby causing observers to underestimate the true luminosity of the \oiii\ $\lambda 5007$ emission line.  Moreover, because higher-mass AGB stars eject a greater amount of material during their evolution, and because their more rapid evolution gives the dust less time to disperse, it is reasonable to expect that PNe from higher-mass progenitors will suffer more circumstellar extinction.  

The new accelerated-evolution models imply that $M^*$ PNe can be produced by progenitors with masses as small as $\sim 1.5 \, \rm{M}_{\odot}$.  This is already a problem for old stellar populations, since $1.5 \, \rm{M}_{\odot}$ stars have main sequence lifetimes of less than 3~Gyr \citep[see, for example,][and references therein]{buzzoni2002}.  However, if circumstellar extinction is important, then this minimum stellar mass must be increased, and, if the extinction is large enough, the \oiii\ $\lambda 5007$ luminosities may completely rule out accelerated evolution as an explanation for the PNLF cutoff.  

Here we address this possibility by measuring the Balmer decrements of 23 of the brightest PNe in the inner bulge ($r < 3\arcmin$) of M31, including seven within $75\arcsec$ of the nucleus, where the light from the classical bulge dominates. We use these data to determine the true minimum \oiii\ $\lambda 5007$ luminosities generated by the $M^*$ PNe and use the \cite{bertolami2016} models to estimate their minimum central star masses.

In Section~2, we discuss details related to our observations and the instrument used to acquire our data.  In Section~3, we describe our reduction and analysis techniques, and in Section 4, we give the circumstellar extinctions and de-reddened magnitudes of the PNe.  Finally,  we discuss the implications of our observations on current models of stellar evolution.

\section{Observations}

The goal of our experiment is to determine the circumstellar extinctions surrounding the \oiii-bright PNe of an unambiguously old stellar population, and to measure the objects' true (de-reddened) \oiii\ $\lambda 5007$ magnitudes.   The best place to do this is in the inner bulge of M31, which is nearby \citep[780~kpc;][]{conn2016}, relatively unaffected by foreground reddening \citep[$A_V \approx 0.24$;][]{conn2016}, and almost devoid of intermediate-age stars \citep{dong2015}.  Moreover, since the region contains $\sim 10^{10} \, \rm{L}_{\odot}$ of starlight, it has a significant PN population, with more than two dozen objects within 1 mag of $M^*$ \citep{ciardullo1989a, ciardullo2002}.

We observe 23 of the region's brightest PNe with the blue feed of the new integral-field unit (IFU) low-resolution spectrograph (LRS2-B) on the Hobby-Eberly Telescope (HET\null).   This double-armed instrument surveys a $12\arcsec \times 6\arcsec$ area on the sky using an array of 280 lenslet-coupled fibers, and delivers a scale of $0\farcs 59$ per fiber. LRS2-B simultaneously observes the spectral range from 3700~\AA\ to 7000~\AA\ with two arms. For this study, we use only the data from the ``Orange Arm'', which produces spectra with a resolution of $R \sim 1100$ between the wavelengths 4600~\AA\ and 7000~\AA\null. We note that because LRS2-B has lenslets, there is no dead space between the fibers, hence many of the problems associated with faint-object spectrophotometry---such as atmospheric dispersion, slit losses, and lack of data acquisition due to imprecise astrometry---are mitigated.   The details and specifications of LRS2 are discussed in more depth in \cite{chonis2016}.

\begin{figure}
    \centering
    \includegraphics[scale=0.61]{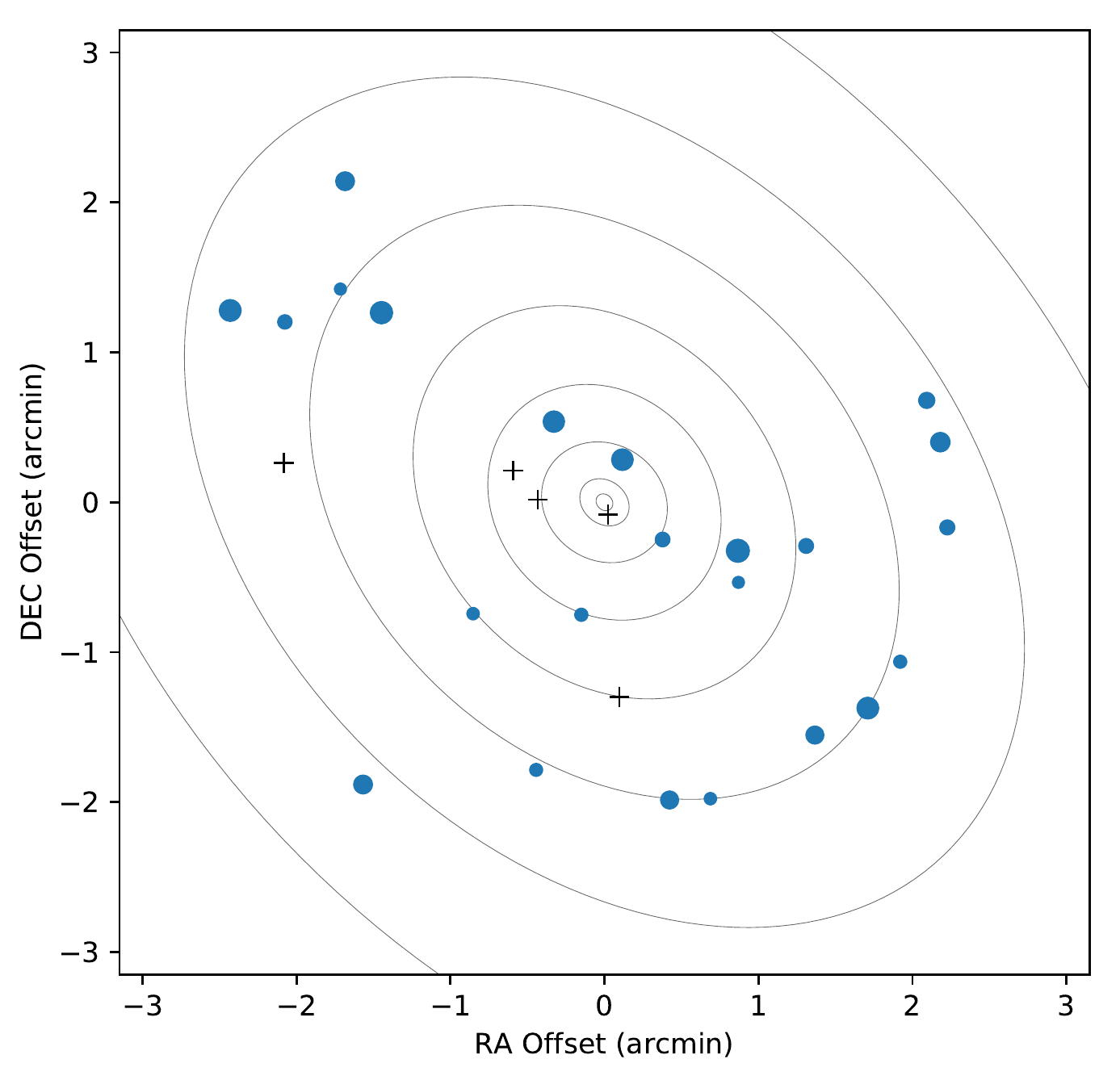}
    \caption{Schematic of M31's inner bulge, with the positions of all PNe with measured \oiii\ $\lambda$5007 fluxes within 1 mag of the observed PNLF cutoff magnitude, $M^*$.  The de-reddened \oiii\ $\lambda$5007 absolute magnitudes of the PNe are reflected by the size of the points, which increase from $M_{5007} = -4.2$ to $M_{5007} = -5.6$. Black crosses denote PNe not observed with the LRS2-B spectrograph.  The contours \citep[from][]{kent1983} represent M31's isophotes (starting at $r = 15.5$~mag~arcsec$^{-2}$ with intervals at 0.5~mag~arcsec$^{-2}$).}
    \label{fig:map}
\end{figure}

We concentrate our observations on the inner $3 \arcmin$ of M31's bulge, and observe 23 of the region's brightest 25 PNe. Table 1 gives a log of these observations, including the original IDs of the PNe \citep{ciardullo1989a}, their IDs from the Sitelle observations of \cite{martin2018}, their J2000 coordinates, the total exposure times (with each observation split in two to facilitate the removal of cosmic rays), and the net image quality for the observation. The locations of these PNe, along those of other PN in the top 1~mag of the PNLF which we did not observe, are shown schematically in Figure~\ref{fig:map}.

\begin{deluxetable*}{cccccccc}
\tablewidth{0pt}
\tablecaption{Log of Observations}
\tablehead{
\colhead{ID} &\colhead{ID} &  & &\colhead{First} &\colhead{Second} &\colhead{Total Exp} &\colhead{Seeing} \\
\colhead{\citep{ciardullo2002}} &\colhead{\citep{martin2018}} &\colhead{$\alpha(J2000)$} &\colhead{$\delta(J2000)$} &\colhead{Observation} &\colhead{Observation} &\colhead{(min)} &\colhead{(arcsec)} }
\startdata
1 & S004246.22+411641.5 & 0 42 46.10 & +41 16 40.9 & Aug 18, 2017 & Nov 24, 2017 & 13.8 & 1.9 \\
3 & S004243.76+411625.9 & 0 42 43.73 & +41 16 25.7 & Sep 14, 2017 & Aug 18, 2017 & 11.4 & 2.1 \\
5 & S004242.38+411553.7 & 0 42 42.34 & +41 15 53.7 & Oct 14, 2017 & Nov 12, 2017 & 15.5 & 2.0 \\
17 & S004239.76+411549.4 & 0 42 39.74 & +41 15 49.2 & Sep 12, 2017 & \dots & 5.2 & 1.9 \\
18 & S004239.76+411536.4 & 0 42 39.72 & +41 15 36.6 & Oct 13, 2017 & \dots & 5.2 & 2.1 \\
21 & S004245.22+411523.4 & 0 42 45.15 & +41 15 23.6 & Oct 17, 2017 & Nov 24, 2017 & 15.5 & 2.3 \\
28 & S004252.13+411724.9 & 0 42 52.06 & +41 17 24.5 & Sep 12, 2017 & \dots & 5.2 & 2.0 \\
29 & S004253.55+411734.3 & 0 42 53.48 & +41 17 33.9 & Oct 17, 2017 & \dots & 5.2 & 1.9 \\
30 & S004255.48+411721.1 & 0 42 55.40 & +41 17 20.8 & Sep 12, 2017 & \dots & 5.2 & 2.1 \\
32 & S004253.40+411817.4 & 0 42 53.32 & +41 18 17.1 & Sep 13, 2017 & \dots & 5.2 & 1.9 \\
36 & S004237.42+411551.1 & 0 42 37.38 & +41 15 51.2 & Oct 13, 2017 & Nov 12, 2017 & 13.0 & 2.1 \\
41 & S004233.22+411649.6 & 0 42 33.21 & +41 16 49.4 & Sep 01, 2017 & \dots & 5.2 & 2.0 \\
42 & S004232.77+411632.9 & 0 42 32.74 & +41 16 32.7 & Sep 01, 2017 & \dots & 5.6 & 2.0 \\
45 & S004232.52+411558.7 & 0 42 32.50 & +41 15 58.5 & Sep 01, 2017 & \dots & 5.6 & 2.1 \\
51 & S004234.16+411504.5 & 0 42 34.13 & +41 15 04.8 & Oct 02, 2017 & \dots & 5.2 & 2.3 \\
53 & S004235.29+411446.0 & 0 42 35.25 & +41 14 46.2 & Sep 01, 2017 & \dots & 5.2 & 2.0 \\
54 & S004237.12+411435.3 & 0 42 37.08 & +41 14 35.5 & Sep 01, 2017 & \dots & 5.9 & 2.0 \\
56 & S004240.72+411409.9 & 0 42 40.69 & +41 14 10.0 & Oct 13, 2017 & \dots & 5.2 & 2.1 \\
57 & S004242.13+411409.3 & 0 42 42.10 & +41 14 09.5 & Oct 14, 2017 & Nov 17, 2017 & 13.0 & 2.0 \\
61 & S004246.75+411421.1 & 0 42 46.71 & +41 14 21.4 & Oct 17, 2017 & \dots & 5.2 & 1.9 \\
62 & S004248.94+411524.0 & 0 42 48.89 & +41 15 24.0 & Oct 17, 2017 & \dots & 5.2 & 2.0 \\
64 & S004252.75+411415.5 & 0 42 52.69 & +41 14 15.7 & Sep 13, 2017 & \dots & 5.2 & 1.9 \\
80 & S004257.37+411725.7 & 0 42 57.30 & +41 17 25.5 & Oct 12, 2017 & Nov 15, 2017 & 14.7 & 2.3 \\
\noalign{\smallskip}
\hline
\noalign{\smallskip}
23 & S004248.99+411656.2 & 0 42 48.89 & +41 16 55.5 & Dec 22, 2017 & \dots & 15.2 & 2.7 \\
27 & S004258.01+411621.5 & 0 42 57.93 & +41 16 21.2 & Dec 22, 2017 & \dots & 10.2 & 2.7 \\
33 & S004257.78+411848.2 & 0 42 57.75 & +41 18 48.3 & Dec 07, 2017 & \dots & 15.2 & 2.2 \\
75 & S004251.39+411556.5 & 0 42 51.31 & +41 15 56.6 & Dec 26, 2017 & \dots & 15.2 & 2.1
\enddata
\tablecomments{Log of observations of bright PNe in the bulge of M31. The first 23 objects listed are the subject of this study, while the last four are observed for external comparison with \cite{jacoby1999}.}
\label{table:spec_obs}
\end{deluxetable*}

\section{Data Reduction}

Our data are initially reduced with the LRS2 Quick-Look Pipeline (\texttt{QLP}; Indahl, B.L., in preparation), an IFU data reduction package specifically designed for the LRS2 spectrographs.  This pipeline automatically finds the correct set of calibration data for each HET observation, re-samples the data both spectrally and spatially, and produces a fully calibrated data cube, which, for the LRS2-B's Orange Arm, is $42 \times 24$ pixels in the spatial directions, and 2,000 pixels in the wavelength direction.  A brief description of this code is presented in the Appendix.

\begin{figure*}
    \centering
    \includegraphics[scale=0.61]{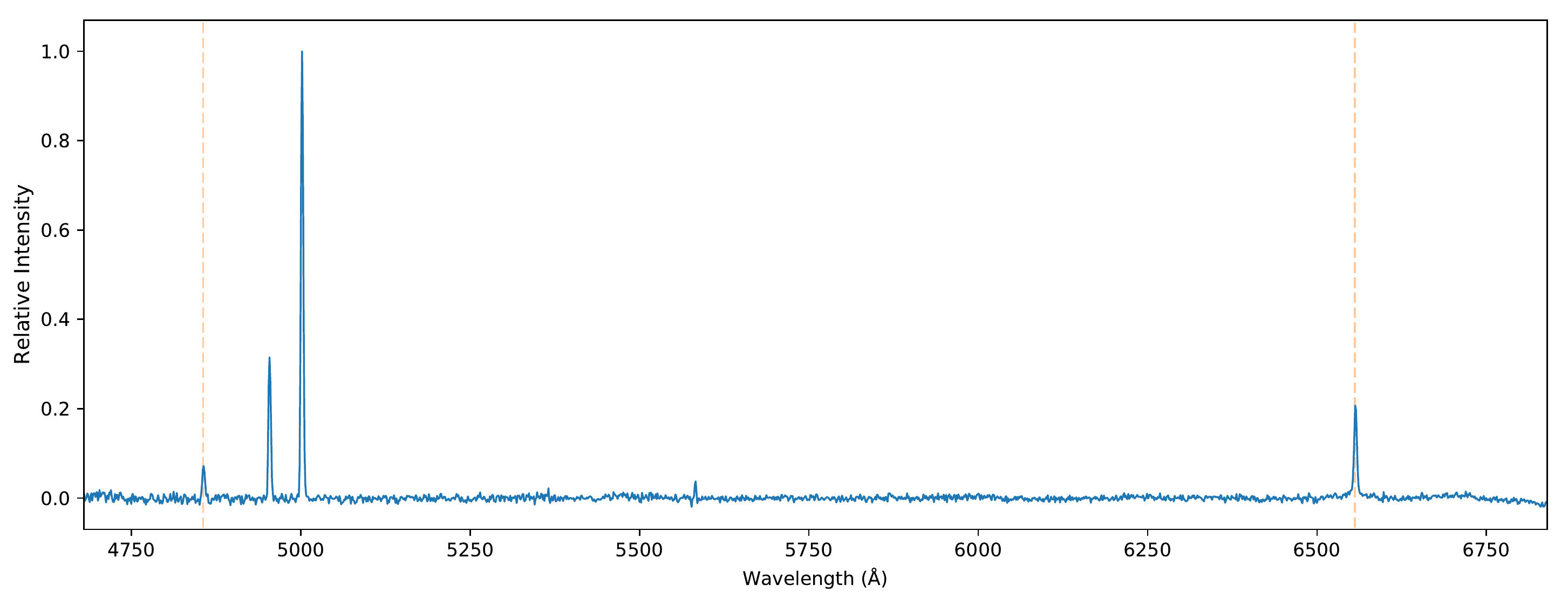}
    \caption{Spectrum from the LRS2-B spectrograph of PN 45 in the bulge of M31. The vertical axis is scaled to the maximum intensity of the \oiii\ $\lambda$5007 line. The faint dashed lines show the locations of H$\beta$ and H$\alpha$, whose flux ratio is used to determine reddening. The PN shown has an apparent \oiii\ $\lambda 5007$ magnitude of $m_{5007} = 21.9$.}
    \label{fig:m31pn45spec}
\end{figure*}

The data cube produced by an observation can be thought of as containing 2,000 unique 2D images, each centered on a different wavelength.  For each PN in our sample, we identify the 2D image that contains the peak of the \oiii\ $\lambda 5007$ line (which depends on the object's Doppler velocity), and use that image to define a circular extraction aperture that is roughly the full width at half maximum (FWHM) of the frame's point spread function (PSF\null).  We then integrate all of the pixels within this aperture, and repeat the procedure in the remaining images. We note that, due to the effects of atmospheric dispersion, the location of the PSF's central peak is not fixed, but instead moves slowly across the image as a function of wavelength. This drift is accounted for and followed by our software. The result is a full 1D spectrum of the targeted PN\null. An example of such a spectrum is given in Figure~\ref{fig:m31pn45spec}. Note that at M31's distance, PNe should have angular sizes that are at least 30 times smaller than that of the typical HET seeing. Thus, issues such as ionization stratification and excitation gradients within the PNe are absolutely no issue for these observations.

The final step in our data reduction is that of flux calibration. To do this, we simply scale the spectrum of each PN to that of a spectrophotometric standard star \citep{stone1977,oke1983} taken on the same night. Because the HET is a fixed-altitude telescope, no airmass corrections are necessary. However, the effective aperture of the telescope may change as objects are tracked across (and sometimes off) the primary's 91 mirror segments. We correct for this effect with a grey term representing the fraction of the HET's primary used during the exposure. Of course, since this study is only concerned with measuring the strength of the H$\beta$ emission line relative to H$\alpha$, this effective aperture term plays no role in our analysis.  

\begin{figure}
    \centering
    \includegraphics[scale=0.61]{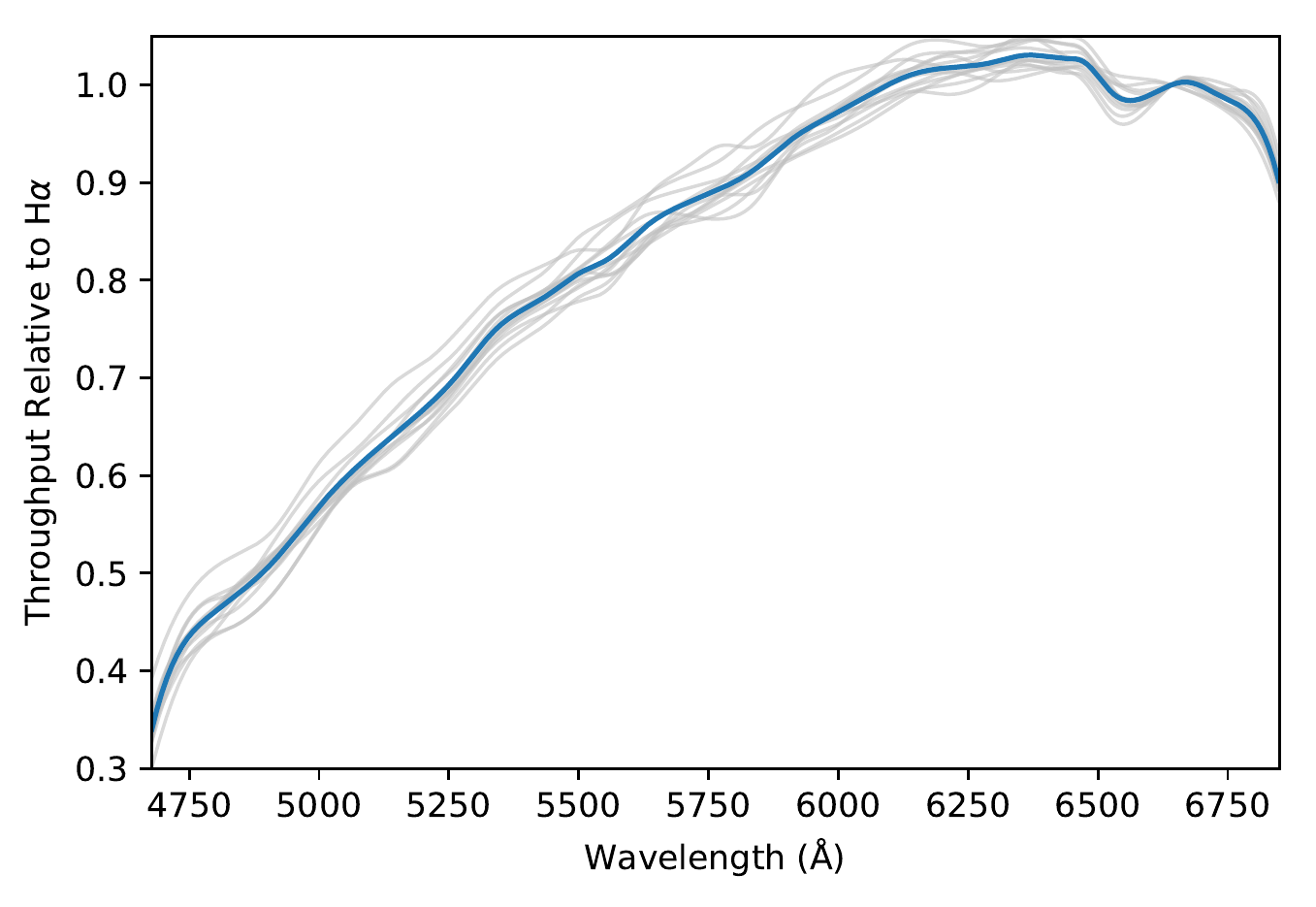}
    \caption{Relative throughput of the Orange Arm of LRS2-B.  All curves have been normalized to 1.0 at H$\alpha$.  The faint lines show the sensitivity on 12 different nights of observations; the bold line displays the median value of the measurements. The throughput at H$\beta$ is generally $0.528 \pm 0.018$ that at H$\alpha$.}
    \label{fig:lrs2B_tp_rel2Ha2}
\end{figure}

\begin{figure}
    \centering
    \includegraphics[scale=0.61]{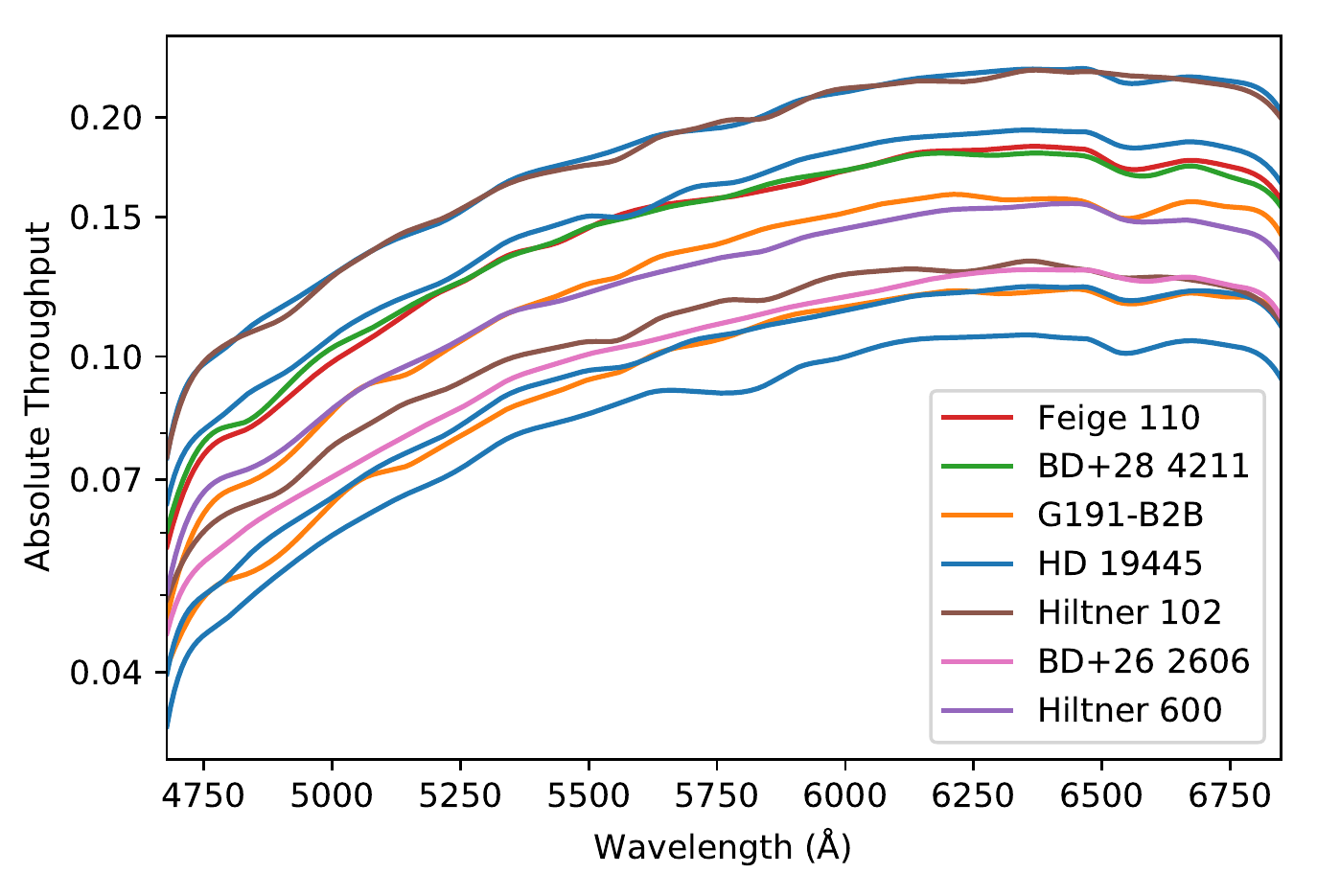}
    \caption{Absolute throughput for the Orange Arm of LRS2-B derived on 12 different nights during the fall of 2017. Each standard star observed is denoted by a different color; some standards are observed more than once. Observations with lower throughput are taken under non-photometric conditions; when conditions are good, the total system throughput between 6000~\AA\ and 6700~\AA\ can be as high as $\sim 20\%$.}
    \label{fig:lrs2B_log_abs_tp}
\end{figure}

Figure~\ref{fig:lrs2B_tp_rel2Ha2} normalizes each of our throughput curves at H$\alpha$ and shows how the sensitivity of the LRS2-B's Orange Arm changes with wavelength. For completeness, Figure~\ref{fig:lrs2B_log_abs_tp} gives the absolute throughput curves (corrected for telescope effective aperture), which represent data from 12 different nights. While not all of our data are taken under photometric conditions, the curves show that the throughput at H$\beta$ relative to H$\alpha$ remains constant to within $\sim 3$\%.

\subsection{M31 Background Subtraction}

For most LRS2-B observations, background subtraction can be performed using \texttt{subtractsky}, a routine called by the \texttt{QLP} pipeline.  However, because our M31 bulge PNe are projected upon a bright, rapidly varying background, the blind use of such automated software can be problematic. Moreover, M31's bulge is known to contain a significant amount of diffuse emission-line gas, whose morphological and kinematic properties are quite complex \citep{jacoby1985, ciardullo1988,melchior2011,opitsch2018}. The removal of this component from PN spectra can be difficult (or impossible)  for single-fiber and long-slit spectrographs. Fortunately, the 2D spatial information provided by the LRS2-B's IFU allows us to address this problem directly.
 
\begin{figure*}
    \centering
    \includegraphics[scale=0.62]{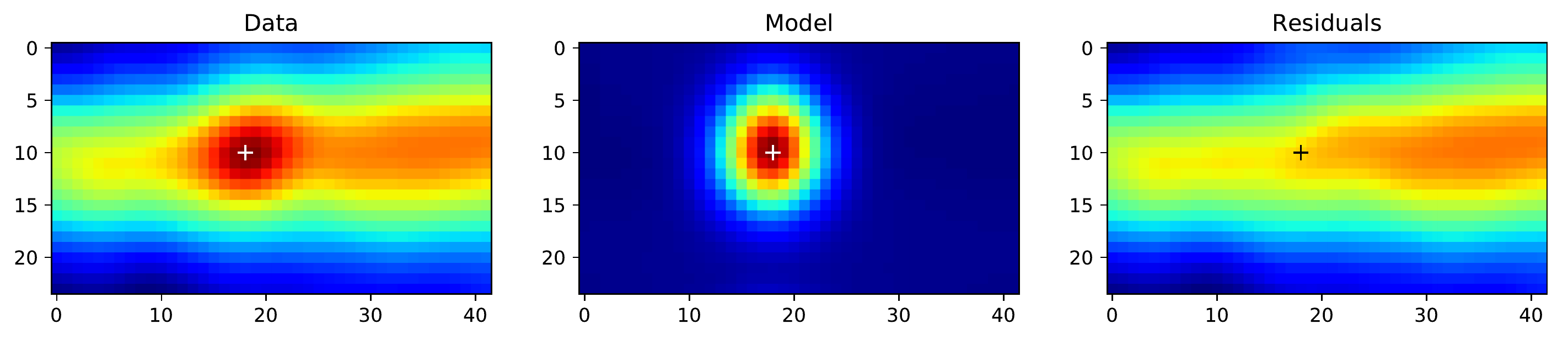}
    \caption{IFU data, model, and background residuals for the H$\alpha$ emission surrounding PN~80 in the bulge of M31.  The axes represent the $(x,y)$ pixel position in the data cube. The color scale is linear and ranges from 7,500 counts (blue) to 11,000 counts (red). Left: the data from the IFU's $12\arcsec \times 6\arcsec$ field-of-view, covering the wavelengths 6544-6568 \AA\null.  Center: a model for the frame's PSF, based on a spectrophotometric standard (HD~19445) observed earlier in the night.  Right: the residual after the subtraction of the scaled PSF. Note that the background appears smooth and continuous. This suggests that our model is a reasonably accurate representation of the PN\null.}
    \label{fig:background_pixels}
\end{figure*}

To properly subtract the diffuse Balmer emission superposed on our objects, we must first obtain an independent estimate of the observation's PSF at H$\alpha$ and H$\beta$.  Because the $12\arcsec \times 6\arcsec$ field of view of our IFU is generally not large enough to contain a bright field star, this is typically done using the observation of the night's spectrophotometric standard. (This can create a problem if the seeing is highly variable over the night. Fortunately, this is not the case during our observations.) We then iteratively scale the PSF until it matches the intensity profile of the PN in question. Once the proper scaling is found, we subtract the model from the data to obtain an image of the PN's underlying diffuse background.  If this background contains discontinuities or otherwise appears unphysical, the process is repeated.  Convergence is achieved when the background appears smooth;  the model then represents the integrated line flux from the PN\null. Figure~\ref{fig:background_pixels} illustrates our procedure.

In the end, only one of our objects (PN 1) requires this full-modeling background removal method. However, the 2D data provided by the IFU is essential for identifying these problem objects. Consequently, we reduce all of the PNe measurements both with and without the \texttt{QLP} pipeline's \texttt{subtractsky} algorithm. A comparison of the results shows that those PNe superposed on locations with bright emission-line gas display Balmer lines with both narrow and broad components.  If the broad components---which tend to be more prominent in H$\alpha$ than in H$\beta$---are removed, then the Balmer flux ratios derived using \texttt{subtractsky} agree to within $\sim 10$\% with those computed using the full background modeling method. This is within the margin of error of our measurements, which gives us some confidence that our Balmer-line flux ratios are robust, even in the presence of a diffuse, irregular background. Any broad Balmer emission present in our spectra is therefore removed from our data before measurement.

\section{Results}

\subsection{Reddening Measurements}

From case B recombination theory, we expect PNe with electron temperatures of $T_e = 10^4$~K and electron densities of $n_e = 100$~cm$^{-3}$ to have Balmer line ratios of $I$(H$\alpha$)/$I$(H$\beta$) = 2.86 \citep{brocklehurst1971, osterbrock2006}.  At the extremes, this ratio can range from 2.74 ($T=20{,}000$~K; $n_e = 10^4$~cm$^{-3}$) to 3.04 ($T=5{,}000$~K; $n_e = 100$~cm$^{-3}$), with the ratio being much more sensitive to temperature than density.  In practice, the electron temperatures for M31 bulge PNe only range from $\sim 8,500$~K to $\sim 16,500$~K \citep{jacoby1999, richer1999}, which implies $I$(H$\alpha$)/$I$(H$\beta$) ratios between 2.78 and 2.90. Here we adopt 2.86 as the intrinsic ratio between these Balmer lines: any value higher than this is assumed to be caused by interstellar (and more importantly, circumstellar) extinction.

To measure the extinction at 5007~\AA, we adopt the \cite{cardelli1989} attenuation law with $R_V = 3.1$  and compute 
\begin{equation}
A_{\lambda} = \frac{2.5 f(\lambda)}{f(\textrm{H}\beta) - f(\textrm{H}\alpha)}\left\{\log\left[\frac{I(\textrm{H}\alpha)}{I(\textrm{H}\beta)}\right] - \log(2.86)\right\},
\end{equation}
where $f(\lambda) = A_{\lambda}/E(B-V)$ are the extinction coefficients (in magnitudes) derived from the relation.  For reference, $f({\rm H}\beta) = 3.61$, $f({\rm H}\alpha) = 2.53$, and $f(\lambda 5007) = 3.47$.  Table \ref{table:A_Vs} gives the derived circumstellar extinctions for our PNe, assuming foreground dust in the Milky Way is responsible for $A_V = 0.24$ of attenuation \citep{conn2016}. Both the measured Balmer ratios and the derived logarithmic extinction at H$\beta$, $c_{{\rm H}\beta}$, are present, along with the PN's isophotal radius, as computed from the surface photometry of \cite{kent1983}.

\begin{deluxetable}{cccccc} 
\tablewidth{0pt}
\tablecaption{Circumstellar Extinctions for PNe in M31's Bulge}
\tablehead{
\colhead{ID\tablenotemark{a}}
& \colhead{$m_{5007}$\tablenotemark{b}}
& \colhead{$m_{\rm{dered}}$\tablenotemark{c}}
& \colhead{H$\alpha$:H$\beta$\tablenotemark{d}}
& \colhead{$c(\mathrm{H}\beta)$\tablenotemark{e}}
& \colhead{$r_i$\tablenotemark{f}} }
\startdata
53 & 20.40 & \textbf{19.27} & 3.95 & 0.36 & 2.19 \\
42 & 20.43 & 19.45 & 3.78 & 0.30 & 2.81 \\
1 & 20.51 & \textbf{19.30} & 4.04 & 0.39 & 0.64 \\
45 & 20.59 & 19.97 & 3.41 & 0.15 & 2.59 \\
54 & 20.61 & 19.60 & 3.81 & 0.31 & 2.08 \\
41 & 20.65 & 19.81 & 3.63 & 0.24 & 2.90 \\
28 & 20.65 & \textbf{19.23} & 4.29 & 0.48 & 1.93 \\
30 & 20.70 & 20.07 & 3.42 & 0.15 & 2.45 \\
17 & 20.71 & \textbf{19.17} & 4.43 & 0.53 & 0.93 \\
64 & 20.78 & 19.51 & 4.10 & 0.42 & 3.56 \\
32 & 20.79 & 19.52 & 4.11 & 0.42 & 2.76 \\
3 & 20.85 & \textbf{19.28} & 4.47 & 0.54 & 0.34 \\
18 & 20.88 & 20.51 & 3.18 & 0.04 & 1.02 \\
61 & 20.93 & 20.32 & 3.41 & 0.15 & 2.39 \\
36 & 20.94 & 19.99 & 3.75 & 0.28 & 1.41 \\
56 & 20.97 & 20.43 & 3.33 & 0.11 & 2.29 \\
80 & 20.98 & \textbf{19.26} & 4.67 & 0.61 & 2.80 \\
29 & 21.01 & 20.51 & 3.29 & 0.10 & 2.24 \\
51 & 21.02 & 20.29 & 3.53 & 0.20 & 2.23 \\
5 & 21.07 & 20.02 & 3.85 & 0.33 & 0.45 \\
62 & 21.09 & 20.42 & 3.46 & 0.17 & 1.45 \\
57 & 21.10 & 19.60 & 4.39 & 0.52 & 2.32 \\
21 & 21.11 & 20.27 & 3.63 & 0.24 & 0.88 \\
\noalign{\smallskip}
\hline
\noalign{\smallskip}
23 & 21.45 & 21.14 & 3.13 & 0.02 & 1.05 \\
27 & 20.85 & 19.62 & 4.06 & 0.40 & 3.04 \\
33 & 21.75 & 20.86 & 3.68 & 0.26 & 3.56 \\
75 & 21.85 & 20.84 & 3.81 & 0.31 & 1.62
\enddata
\tablecomments{M31 bulge PN circumstellar extinctions assuming a foreground Milky Way component of $A_V = 0.24$ \citep{conn2016}.  The six brightest de-reddened \oiii\ $\lambda$5007 magnitudes are shown in bold. The first 23 objects listed are the subject of this study, while the last four are observed for external comparison with \cite{jacoby1999}.}
\tablenotetext{a}{PNe numbered as in \cite{ciardullo1989a}.}
\tablenotetext{b}{Observed \oiii\ $\lambda 5007$ magnitude from Ciardullo \etal (1989a, 2002).}
\tablenotetext{c}{De-reddened \oiii\ $\lambda 5007$ magnitude.}
\tablenotetext{d}{Observed Balmer flux ratio.}
\tablenotetext{e}{Circumstellar logarithmic extinction at H$\beta$ (after removing foreground extinction).}
\tablenotetext{f}{Isophotal distance from the center of M31 (in arcmin). Computed from surface photometry of \cite{kent1983}.}
\label{table:A_Vs}
\end{deluxetable}
	
\begin{figure}
    \centering
    \includegraphics[scale=0.62]{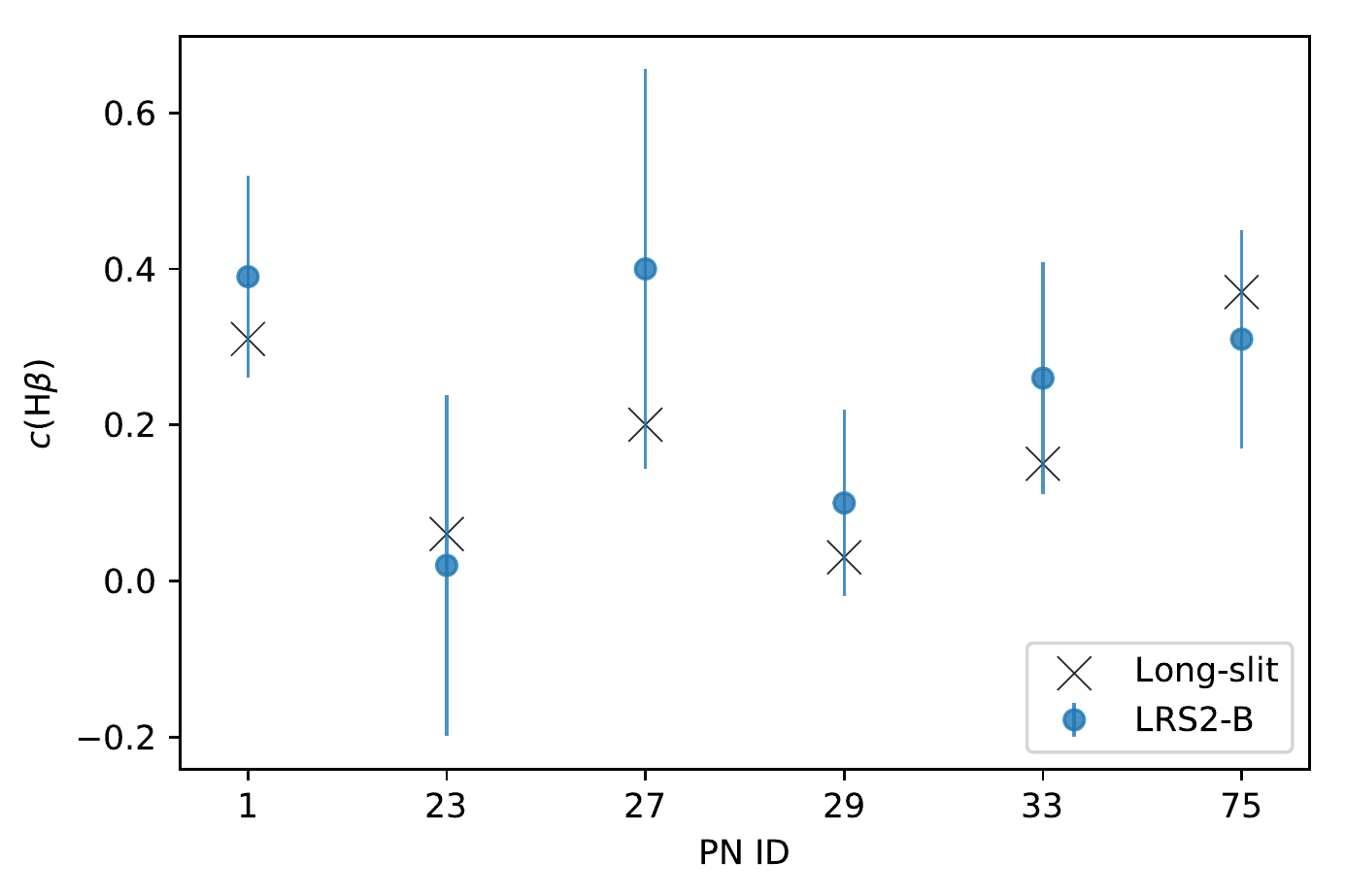}
    \caption{Our LRS2-B circumstellar $c($H$\beta)$ extinctions compared with those derived by \cite{jacoby1999} from long-slit spectroscopy.  The LRS2-B measurements are shown with blue circles and error bars; the \cite{jacoby1999} data are shown with black crosses. The comparison illustrates that the two surveys produce consistent reddenings.}
    \label{fig:ifu_vs_ls}
\end{figure}

\begin{figure}
    \centering
    \includegraphics[scale=0.62]{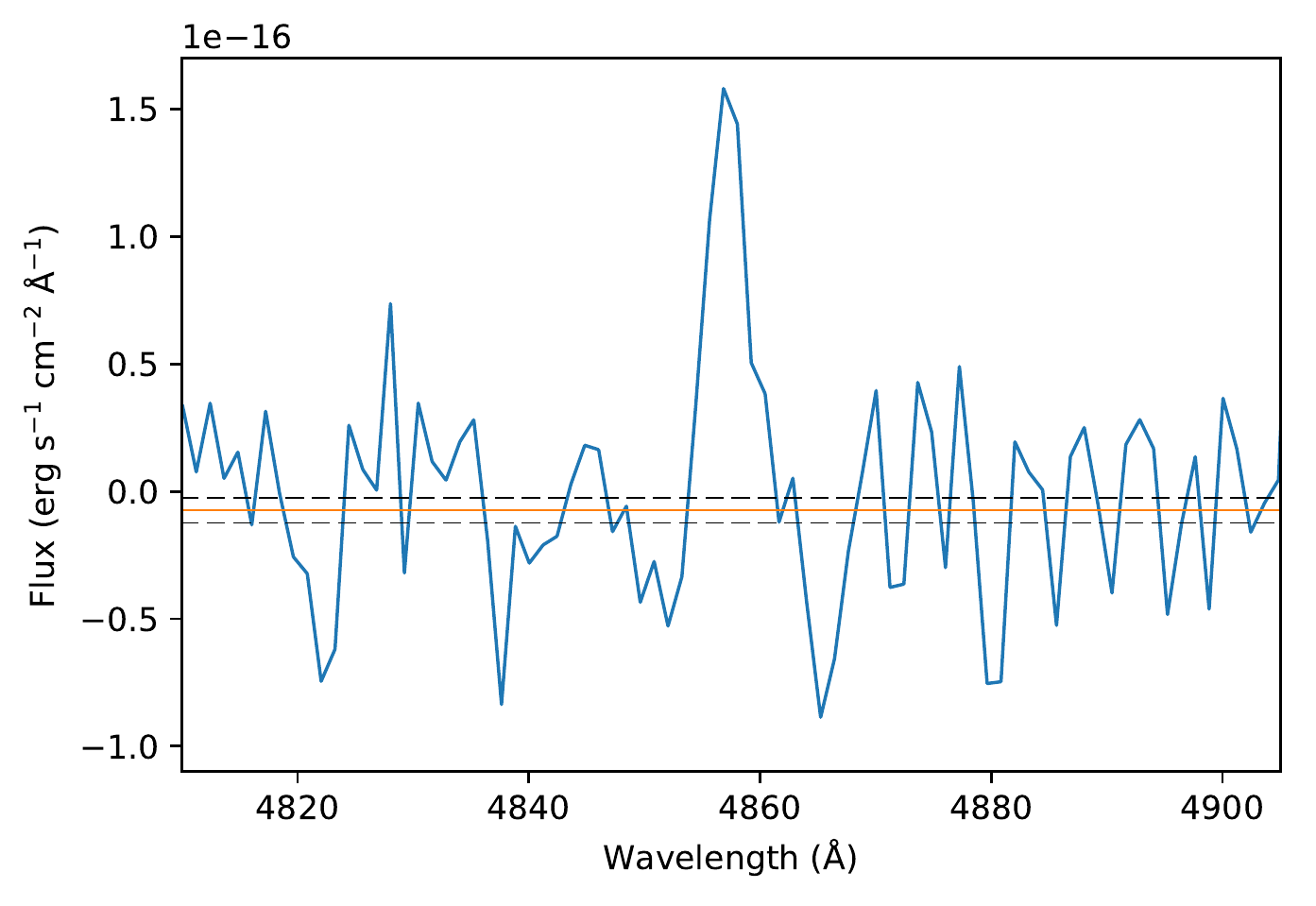}
    \caption{The H$\beta$ emission line of PN 80, after background subtraction.  Our estimate of the continuum, as defined by the mean value of the spectrum between 4800~\AA\ and 4900~\AA\ (excluding the H$\beta$ line itself) is shown by the solid orange line.  The dashed lines show the standard deviation of this mean.  In general, the uncertainty in the H$\beta$ continuum measurement dominates our error budget.}
    \label{fig:Hb_cont_pn80}
\end{figure}

To examine the precision of our H$\alpha$ to H$\beta$ line ratios, we use the observations of \cite{jacoby1999}, who measured Balmer decrements for 12 of M31's bulge PNe via long-slit spectroscopy.  Since only two of the objects from this dataset, PN 1 and PN 29, are part of our sample of bright PNe, we supplement our data by observing four fainter PNe explicitly for the purpose of an external comparison.  As Figure~\ref{fig:ifu_vs_ls} illustrates, our $c(\rm{H}\beta)$ measurements are in close agreement with those of \cite{jacoby1999}.  The standard deviation between the two sets of measurements is $\sigma_{\rm ext} = 0.089$; for comparison, if we assume that our major sources of uncertainty come from setting the PN continua (see Figure~\ref{fig:Hb_cont_pn80}) and the scatter in instrument's response function (Figure~\ref{fig:lrs2B_tp_rel2Ha2}), then the internal error in our measurements is $\sigma_{\rm int} = 0.106$. As Table~\ref{table:error} illustrates, other sources of error are minor in comparison. Thus, within the known observational errors, the two sets of data appear consistent.

\begin{deluxetable}{llc} 
\tablewidth{0pt}
\tablecaption{$M_{5007}$ Error Budget}
\tablehead{
\colhead{Type}
& \colhead{Source}
& \colhead{Error ($M_{5007}$)} }
\startdata
Random & Relative throughput\tablenotemark{a} & 0.12 \\
& Spectral continuum\tablenotemark{b} & 0.26 \\
& Intrinsic Balmer ratio\tablenotemark{c} & 0.06 \\
& $m_{5007}$\tablenotemark{d} & 0.05 \\
\noalign{\smallskip}
\cline{2-3}
\noalign{\smallskip}
& All random error & 0.30 \\
\noalign{\smallskip}
\hline
\noalign{\smallskip}
Systematic & Extinction law\tablenotemark{e} & 0.04 \\
& M31 distance\tablenotemark{f} & 0.05 \\
\noalign{\smallskip}
\cline{2-3}
\noalign{\smallskip}
& All systematic error & 0.06 \\
\noalign{\smallskip}
\hline
\noalign{\smallskip}
Both & All error & 0.30
\enddata
\tablecomments{Sources of error in the \oiii\ $\lambda$5007 absolute magnitude measurement for typical PNe in our sample.}
\tablenotetext{a}{Defined from the spread of the LRS2-B throughput at H$\beta$ relative to H$\alpha$ (see Figure~\ref{fig:lrs2B_tp_rel2Ha2}).}
\tablenotetext{b}{Based on the uncertainty in our measurement of the continuum underlying the emission lines (see 
Figure~\ref{fig:Hb_cont_pn80}).}
\tablenotetext{c}{From the choice of $[I(\textrm{H}\alpha)/I(\textrm{H}\beta)]_0 = 2.86$, assuming the ratios can vary between 2.78 and 2.90.}
\tablenotetext{d}{From \cite{ciardullo2002}.}
\tablenotetext{e}{Estimated by comparing the \cite{cardelli1989} extinction law to those from \cite{savage1979}, \cite{seaton1979}, and \cite{fitzpatrick1999}.}
\tablenotetext{f}{From \cite{conn2016}.}
\label{table:error}
\end{deluxetable}

\subsection{Distribution of Circumstellar Extinctions}

\begin{figure}
    \centering
    \includegraphics[scale=0.63]{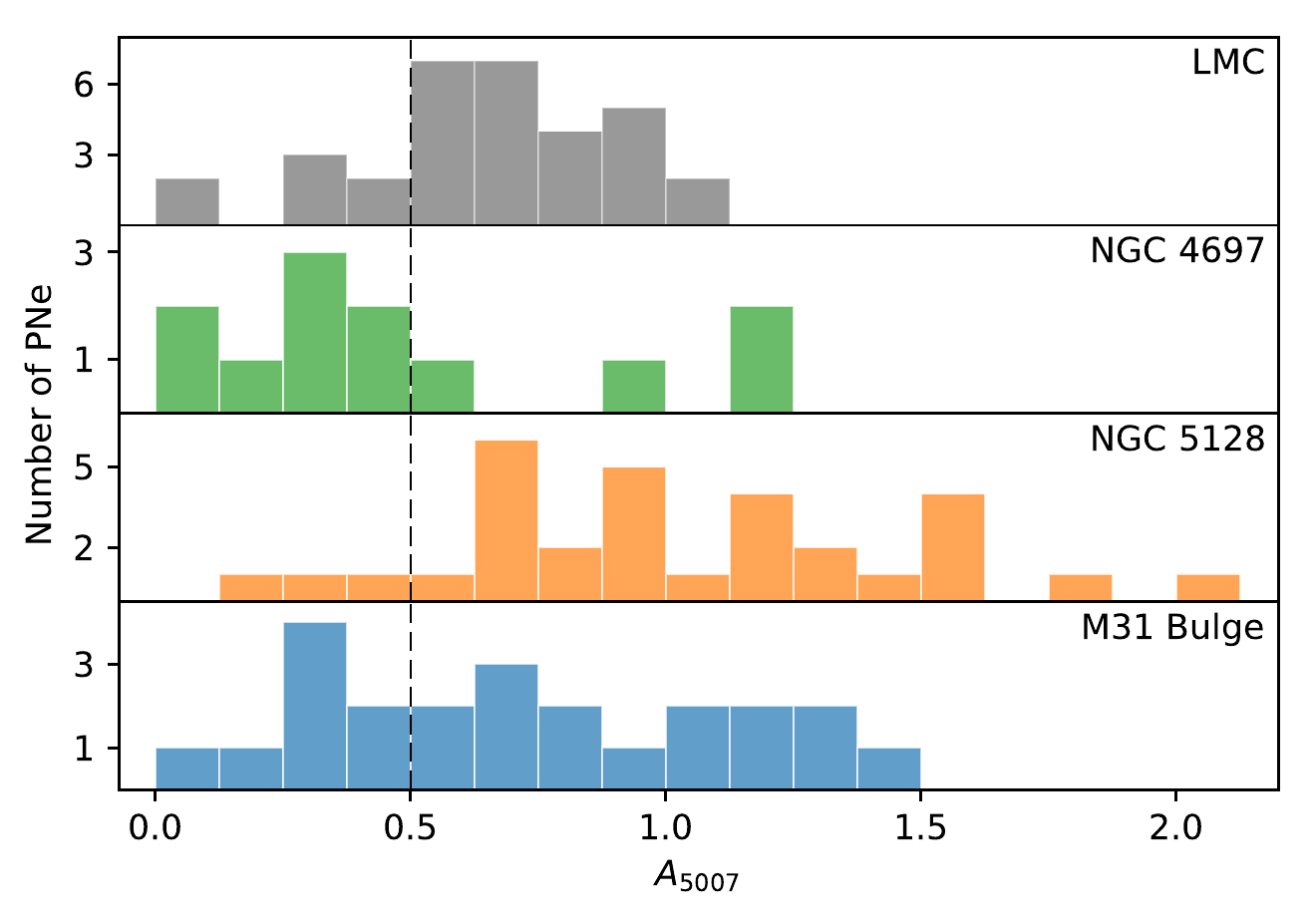}
    \caption{Distributions of circumstellar \oiii\ $\lambda 5007$ extinctions of the brightest PNe in the LMC \citep[30 objects from][]{reid2010}, NGC~4697 \citep[14 objects from][]{mendez2005}, NGC~5128 \citep[44 objects from][]{walsh2012}, and the bulge of M31 (23 objects). All samples consist of PNe within the brightest 2 mag of the de-reddened PNLF (the brightest $\sim 1$ mag of the observed PNLF\null). The dashed line shows the value of  $A_{5007}=0.5$ extinction assumed by \cite{gesicki2018} in their analysis.  Most of the brightest PNe have circumstellar extinctions larger than this.}
    \label{fig:all_cHb_hist}
\end{figure}

The bottom panel of Figure~\ref{fig:all_cHb_hist} displays the distribution of circumstellar extinctions surrounding the bright PNe in the bulge of M31.  As the figure illustrates, the self-attenuation associated with these PNe is substantial.  In the median, the observed \oiii\  $\lambda 5007$ magnitudes are affected by $A_{5007} = 0.71$~mag. There is a strong correlation between intrinsic \oiii\ $\lambda 5007$ magnitude and attenuation, but this is likely a selection effect: because the PNe in our sample are all drawn from the brightest $\sim 1$~mag of the observed PNLF, the intrinsically brightest PNe will generally be the objects with the largest reddening correction. As a rule, the objects that define the bulge's PNLF cutoff are roughly twice as bright as they appear.

To compare our results with those of other PN samples, we begin with dataset of \cite{reid2010}, who have measured hundreds of reddenings and absolute magnitudes in the star-forming population of the Large Magellanic Cloud (LMC\null).  Despite having a mean metallicity significantly less than that of M31's bulge \citep{richer1993, jacoby1999}, the PNe of the LMC display a similar distribution of circumstellar extinctions, with a peak near $A_{5007} \approx 0.6$~mag, and a median value of $A_{5007} = 0.71$. If we restrict the sample to only those PNe within the brightest 2 mag of the de-reddened PNLF, the median extinction decreases somewhat to $A_{5007} = 0.64$, and it is this distribution that is plotted in the top panel of Figure~\ref{fig:all_cHb_hist}.   Both these distributions are consistent with the mean value of $A_{5007} \approx 0.7$ obtained by \cite{herrmann2009} from the co-added spectra of hundreds of bright PNe in the disks of nearby spirals.

For an older, intermediate-age population, we can examine the Balmer ratios of PNe in the interacting lenticular galaxy NGC~5128. At a PNLF distance of $\sim 3.4$~Mpc \citep{ciardullo2002}, the system is our nearest large, early-type galaxy. Although its envelope has a color more akin to the interarm region of a spiral galaxy, rather than a true elliptical \citep{dufour1979}, the galaxy's color-magnitude diagram \citep{rejkuba2011} and the faint-end behavior of its PNLF \citep{ciardullo2010} both suggest an old stellar population that is quite different from that of the LMC\null. \cite{walsh2012} performed spectrophotometry on more than 50 of the galaxy's bright PNe (within $\sim 2$~mag of $M^*$), obtaining reliable reddenings for 44 of the objects. Although some of the PNe may be affected by the galaxy's own (very patchy) internal reddening, we see that, once the foreground Milky Way extinction of  $A_V = 0.315$ \citep{schlafly2011} is removed, the overall extinction distribution is again similar to that seen in M31.  The sample's median attenuation is $A_{5007} =  0.89$.  

Unfortunately, it is difficult to measure PN extinctions in a true elliptical galaxy since the nearest of these systems is $\sim 10$~Mpc away.  Nevertheless, by combining Keck and VLT data, \cite{mendez2005} were able to obtain spectrophotometry for 14 PNe in the normal E6 elliptical NGC~4697 \citep[$D_{\rm{PNLF}} \approx 9.5$~Mpc;][]{ciardullo2002}. This is a very small sample, and any attempt at quantitative analysis will likely be compromised by strong selection effects.  Nevertheless, when the Milky Way's foreground extinction of $A_V = 0.081$ \citep{schlafly2011} is removed, the distribution of circumstellar extinctions has a median value of $A_{5007} = 0.29$. A list summarizing the distribution of PN circumstellar extinctions in each of these galaxies is given in Table \ref{table:dists}.

\begin{deluxetable}{lcccc} 
\tablewidth{0pt}
\tablecaption{Distributions of Circumstellar $A_{5007}$ Extinctions}
\tablehead{
\colhead{Population}
& \colhead{$N$\tablenotemark{a}}
& \colhead{Median}
& \colhead{Mean}
& \colhead{Dispersion} }
\startdata
LMC & 572 & 0.71 & 0.77 & 0.56 \\
LMC (Bright End) & 35 & 0.64 & 0.57 & 0.33 \\
NGC 4697 & 14 & 0.29 & 0.36 & 0.47 \\
NGC 5128 & 44 & 0.89 & 0.90 & 0.47 \\
M31 Bulge & 23 & 0.71 & 0.73 & 0.38
\enddata
\tablenotetext{a}{Number of PNe in the sample.}
\label{table:dists}
\end{deluxetable}

As Figure~\ref{fig:all_cHb_hist} illustrates, the bright bulge PNe of M31 have circumstellar extinctions that are, on average, larger than those associated with the  PNe of NGC~4697 and the LMC, but smaller than those found in NGC~5128.  Since the metallicity of these bulge objects likely falls between that of the LMC  \citep{sarajedini2005, choudhury2016} and NGC~5128 \citep{rejkuba2011}, this result seems to agree with that of \cite{stanghellini2012}, who suggest higher metallicity populations tend to  produce dustier PNe. However, a larger sample of objects is needed to confirm this trend.

\subsection{The De-reddened PNLF}

\begin{figure}
    \centering
    \includegraphics[scale=0.64]{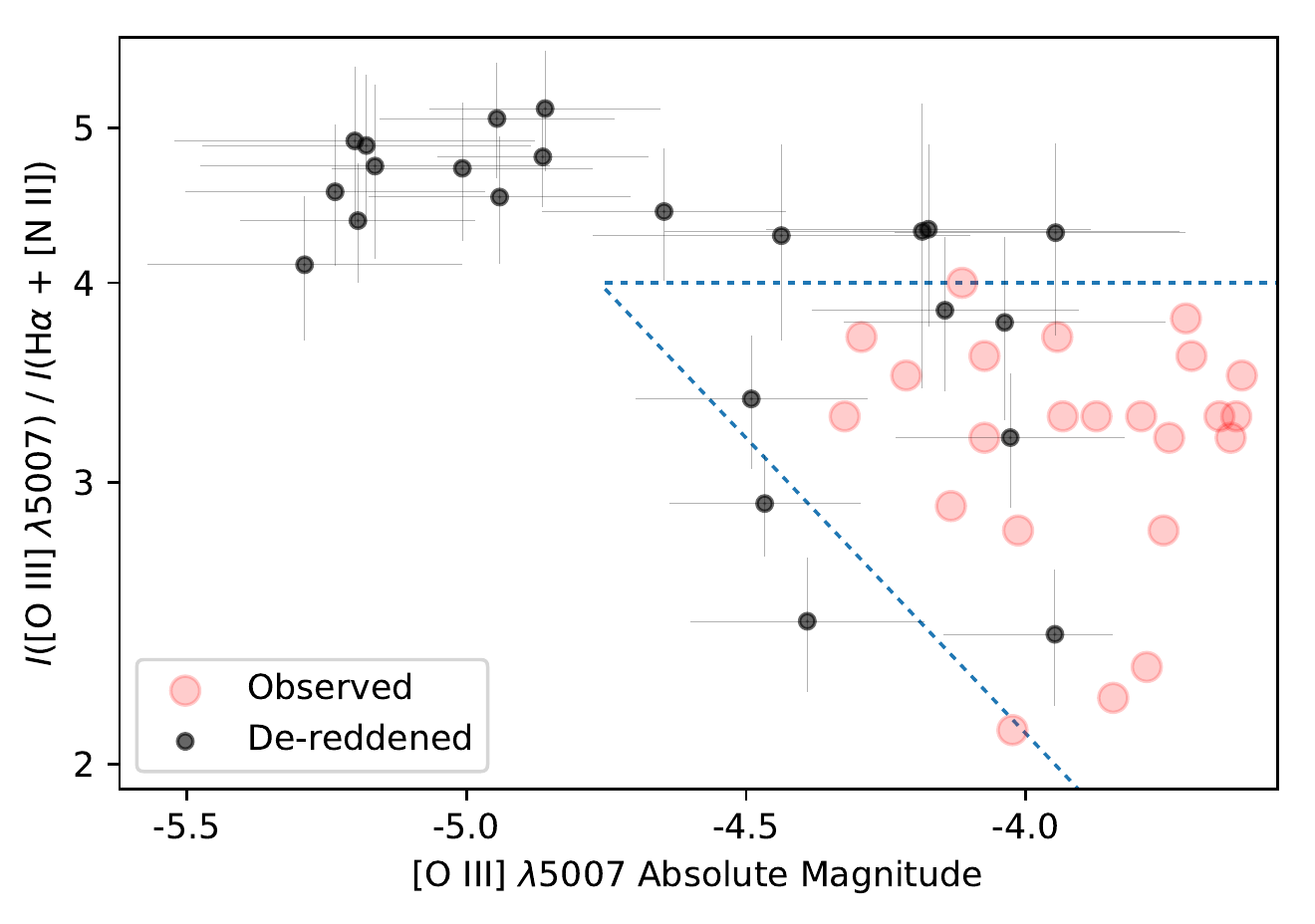}
    \caption{Bright PNe in the bulge of M31, before and after de-reddening.  The vertical axis is the ratio of the \oiii\ $\lambda 5007$ flux to the H$\alpha$+\nii\ flux, as recorded in narrow-band images; the errors in this ratio are correlated with our errors in \oiii\ $\lambda 5007$ absolute magnitude (horizontal axis). The red circles and black points represent the observed and de-reddened PNe, respectively. The PN cone, defined by \cite{herrmann2008}, is shown by the dotted blue lines in the lower right region of the plot.  The brightest de-reddened PNe are up to $\sim 1$~mag more luminous than previously assumed.}
    \label{fig:dered_cone}
\end{figure}

\begin{figure}
    \centering
    \includegraphics[scale=0.63]{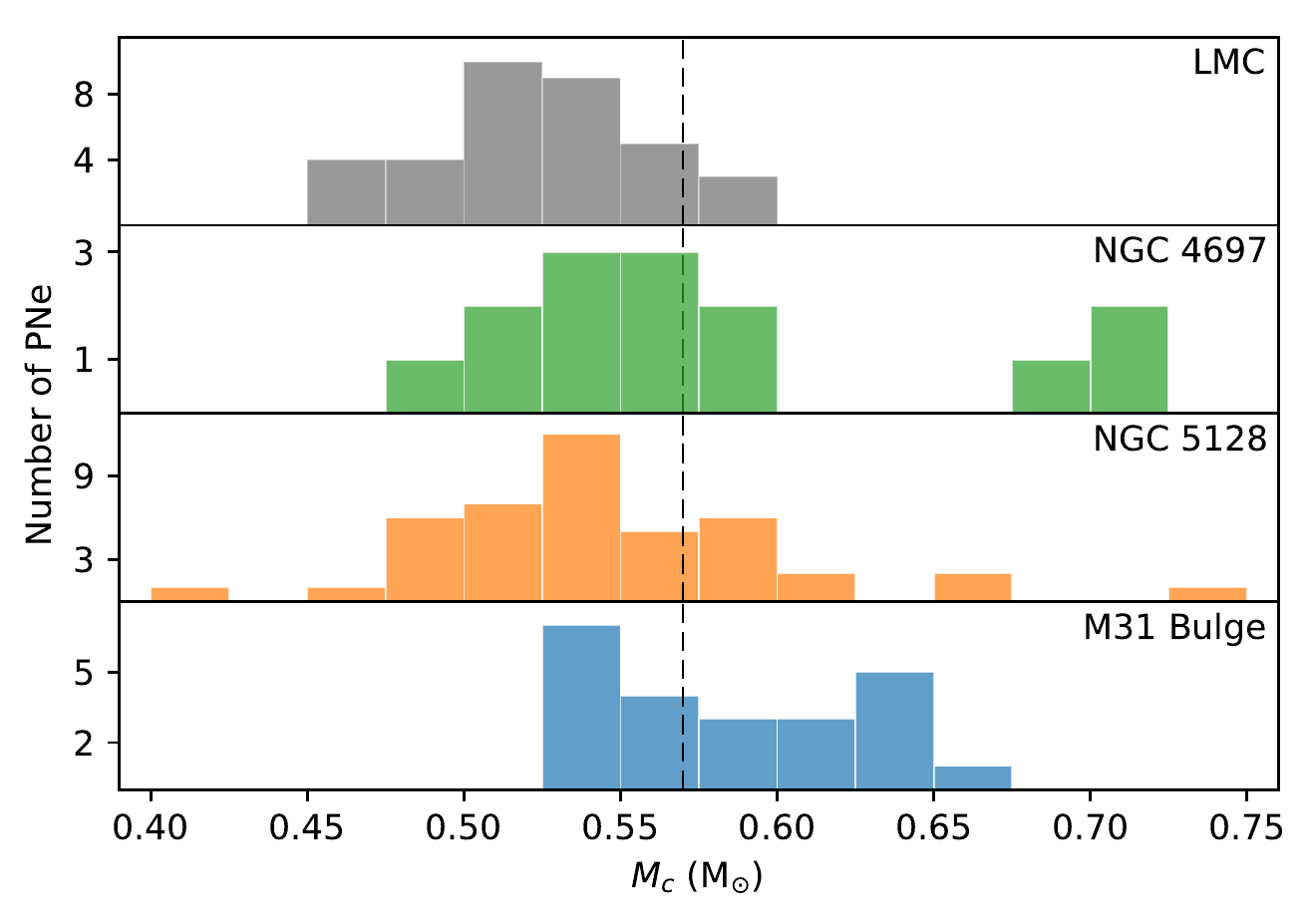}
    \caption{Inferred minimum central star masses for PNe in the LMC, NGC~4697, NGC~5128, and the bulge of M31, as derived from the accelerated post-AGB models of \cite{bertolami2016}. The dashed line at $M_c = 0.57 \, \rm{M}_{\odot}$ gives the minimum central star mass that can achieve a PN absolute magnitude of $M^* = -4.54$. The PN central stars must be more massive than this to explain their de-reddened magnitudes.}
    \label{fig:all_Mc_hist}
\end{figure}

\begin{figure}
    \centering
    \includegraphics[scale=0.63]{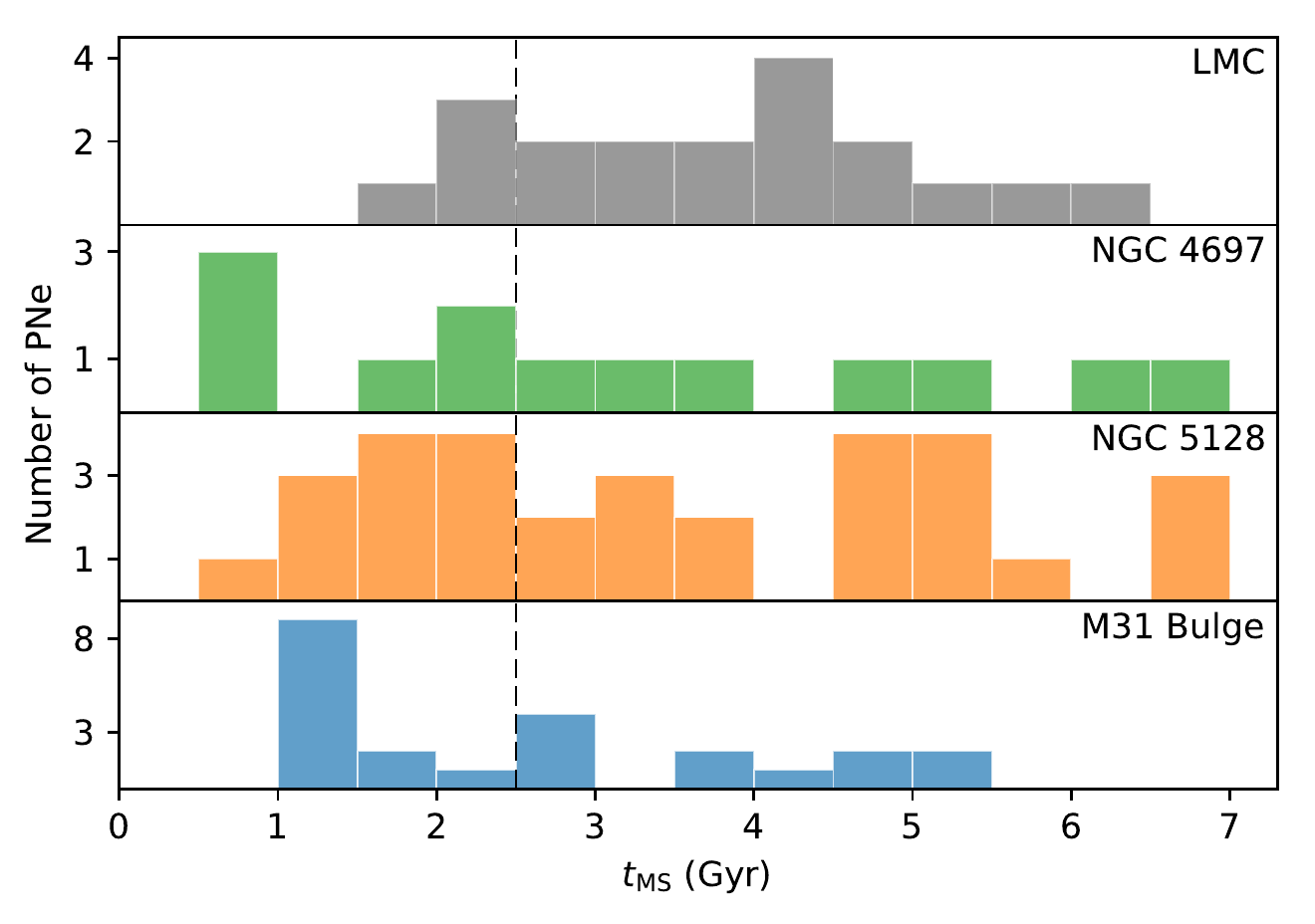}
    \caption{Inferred maximum main sequence lifetimes for the progenitors of PNe in the LMC, NGC~4697, NGC~5128, and the bulge of M31, as derived from the accelerated post-AGB models of \cite{bertolami2016}. The dashed line at $t_{\rm MS} = 2.5$~Gyr gives the main sequence lifetime of the progenitor of a PN with a central star mass of $0.57 \, \rm{M}_{\odot}$, as per initial-final mass relation of \cite{kalirai2008}. Stars with such short main sequence lifetimes are not expected to be found in old populations.}
    \label{fig:all_tMS_hist}
\end{figure}

The reddening distributions shown in Figure~\ref{fig:all_cHb_hist} demonstrate that the amount of \oiii\ $\lambda 5007$ emission produced by an $M^*$ planetary is significantly larger than $\sim 600 \, \rm{L}_{\odot}$.  In fact, as illustrated in Figure~\ref{fig:dered_cone}, the \oiii\ $\lambda 5007$ luminosities of the 23 bright PNe in M31's bulge can be up to $\sim 1$ mag more luminous than the canonical value of $M^*$.  If we assume that \oiii\ $\lambda 5007$ is produced with a maximum efficiency of 11\% \citep{dopita1992, schonberner2010, gesicki2018}, then the central stars of these objects must be brighter than $11{,}000\, \rm{L}_{\odot}$. Even if we adopt the new, accelerated-evolution post-AGB models of \cite{bertolami2016},  these luminosities require central star masses in excess of $\sim 0.66 \, \rm{M}_{\odot}$, meaning that if the initial-final mass relation holds true, their progenitors must have had masses $\gtrsim 2.5 \, \rm{M}_{\odot}$ \citep{kalirai2008, casewell2015}. Such intermediate-mass stars have lifetimes less than 1 Gyr! These minimum central star masses and progenitor lifetimes are shown in Figures \ref{fig:all_Mc_hist} and \ref{fig:all_tMS_hist}. 

Do such stars exist in the bulge of M31?  \cite{dong2015} found that intermediate-age stars likely represent $\sim 1$\% of the total stellar mass in the galaxy's inner bulge ($r <3\arcmin$).   To test whether such a small component can account for the high-mass PN central stars that we observe, we can estimate the number of PNe we might expect to observe in such a population.  The bolometric luminosity-specific stellar evolutionary flux for PN-forming stellar populations is roughly $b = 2 \times 10^{-11}$~yr$^{-1}~\rm{L}_{\odot}^{-1}$, independent of age, metallicity, and initial mass function \citep{renzini1986}.  If we integrate the total amount of optical stellar luminosity contained within M31's central $3\arcmin$ \citep{deVaucouleurs1958, kent1983}, apply a bolometric correction of $-0.85$ \citep{buzzoni2006}, and assume that $\sim 1\%$ of this light comes from intermediate-mass stars, then the total bolometric luminosity of the intermediate-age stars in the bulge of M31 is  $L \approx 10^8 \, \rm{L}_{\odot}$.   Finally, \cite{gesicki2018} estimate that \oiii-bright PNe can spend at most $t \approx 10^3$~yr at luminosities brighter than $M^*$.  Thus, we would expect no more than $b \, t L \approx 2$~PNe as bright as $M^*$ to arise from the inner bulge's intermediate-age population. We observe more than 10 of these very bright PNe within the inner bulge. While the \cite{dong2015} estimate of the intermediate-age population may be uncertain by up to a factor of two, in no way can it produce the observed population of \oiii-bright PNe.

\begin{figure}
    \centering
    \includegraphics[scale=0.63]{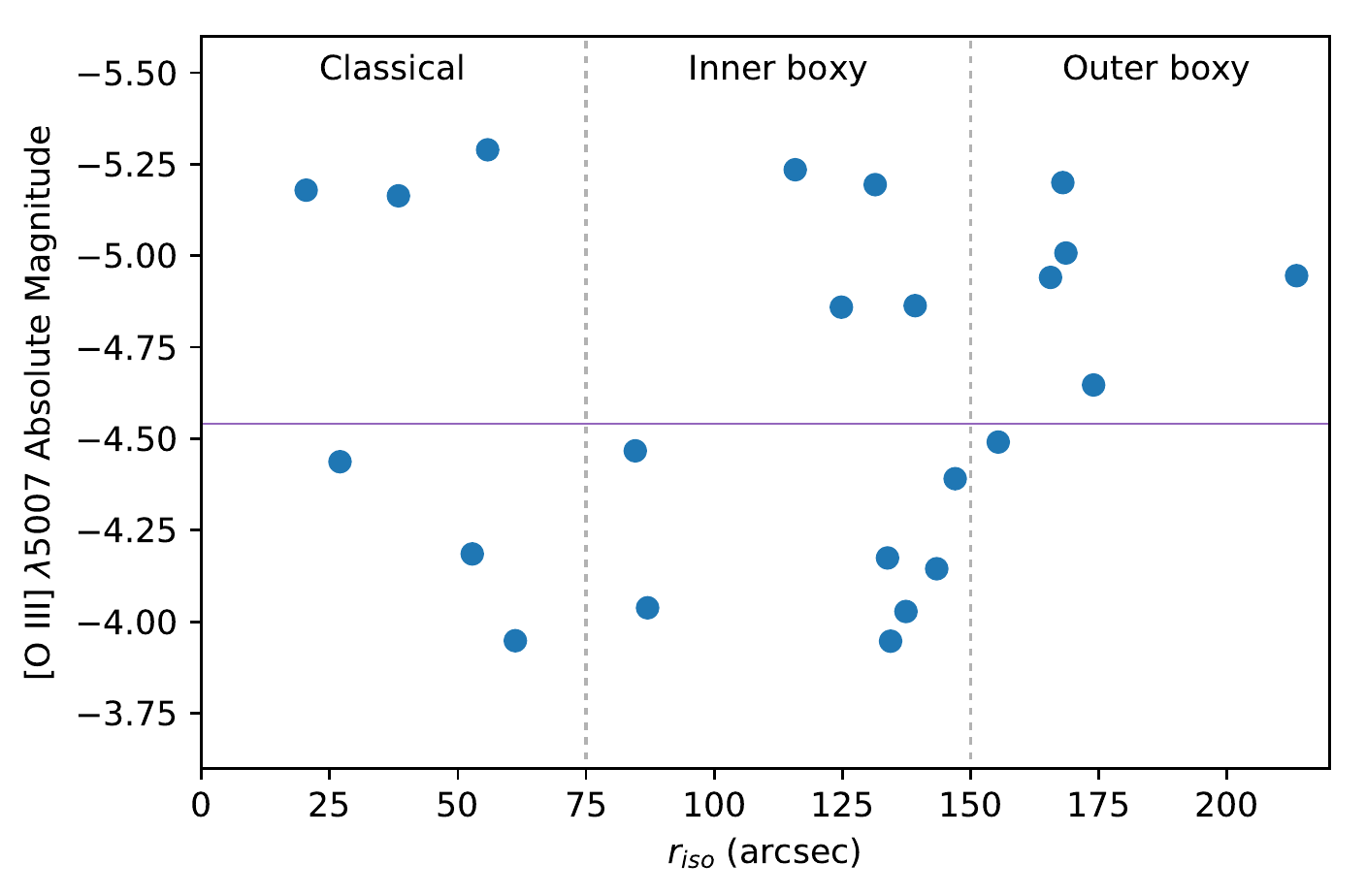}
    \caption{The de-reddened absolute $M_{5007}$ magnitudes of our M31 bulge PNe plotted against their isophotal distance from the center of the galaxy.  The horizontal solid line denotes the observed bright-end cutoff of the PNLF, $M_{5007} = -4.54$. The vertical dashed lines denote the different bulge components, as modeled by \citep{blana2017}. Bright, high-mass PN central stars are found in both the inner classical bulge, and the more extended box/peanut component.}
    \label{fig:M5007vsR}
\end{figure}

Alternatively, since the \cite{dong2015} result is less robust at galactocentric distances greater than $100\arcsec$ (due to missing data in the F390M, F547M and F665N bands), we could assume that their estimate of $\sim 1$\% for the intermediate age population only refers to M31's inner bulge, and that the region outside $\sim 0.4$~kpc has a different stellar population. Indeed, \cite{blana2017} have developed a model in which the light from M31's compact classical bulge gives way to a more extended, box/peanut component, whose contribution only becomes significant at distances greater than $\sim 2\arcmin$.  However, this dynamical decomposition still cannot explain our PN observations. First, as Figure~\ref{fig:M5007vsR} illustrates, the inner $75\arcsec$ of M31's bulge contains three extremely luminous PNe (with de-reddened absolute magnitudes $M_{5007} < -5.1$); according to the \cite{dong2015} SED-fitting analysis, the region's intermediate-age population should produce less than one such object. Second, there is little evidence for intermediate age stars in M31's non-classical bulge. According to \cite{blana2017}, the mass-to-light ratio of M31's box/peanut component is $\Upsilon_{3.6~\mu{\rm m}} \sim 0.81 \rm{M}_{\odot}/\rm{L}_{\odot}$. This estimate, coupled with the region's red color \citep[$B-R \sim 1.6$;][]{tempel2010}, implies a population that is predominantly old: when interpreted using an exponentially declining star formation rate history, the best-fit value for the e-folding rate is only $\tau \sim 2$~Gyr \citep{meidt2014}. In other words, the percentage of intermediate age stars associated with the component should be similar to that estimated by \cite{dong2015}, i.e., $< 2$\%, and the region should contain much less than one PN from an intermediate age progenitor. However, we observe eight PNe brighter than $M^*$ at distances between 75\arcsec and 180\arcsec. The similarity between the population of M31's classical bulge and box/peanut component is also apparent from the region's PN population: as Figure~\ref{fig:M5007vsR} illustrates, high-mass PN central stars are not associated with any single dynamical component, but are distributed throughout the inner regions of the galaxy.

These arguments demonstrate the difficulty associated with explaining the PNLF's bright-end cutoff using conventional single-star stellar evolution.  Hydrogen fusion generates $\sim 6 \times 10^{18}$~erg~g$^{-1}$. Accelerated-evolution models, such as those by \cite{gesicki2018} may generate higher-luminosity post-AGB central stars by burning the hydrogen faster, but they also reduce the stars' evolutionary timescales.  In other words, the parameter space available for PNLF models is defined not only by the absolute luminosity of these objects, but the number of objects as well.  It is extremely difficult, if not impossible, for single-star evolution in old stellar populations to satisfy both constraints.

The de-reddened absolute magnitudes in NGC~4697 and NGC~5128 demonstrate that M31's bulge is not unique.  Both of these systems also contain extremely luminous PNe, with maximum absolute magnitudes of $-5.6$ and $-5.7$, respectively. If we again assume a maximum efficiency of stellar luminosity to \oiii\ $\lambda 5007$ emission of 11\%, and use the \cite{bertolami2016} models as a guide, then the central stars of these extremely luminous PNe must be at least $0.72 \, \rm{M}_{\odot}$.  While one might expect such central stars in open clusters such as Praesepe, these objects should not be present in old, red populations.

We note that any underestimate of the distances to these galaxies will lead to an underestimate of the absolute magnitudes---and thus the central star masses---of these brightest PNe.  So, if the suggestion of \cite{ferrarese2000} is correct, and PNLF distances beyond $\sim 10$~Mpc are underestimated by up to $\sim 0.2$~mag, then the de-reddened absolute magnitude of the brightest PN in NGC~4697 could be as bright as $-5.9$. This would imply the existence of central stars with masses of at least 0.79~M$_{\odot}$.

\begin{figure}
    \centering
    \includegraphics[scale=0.64]{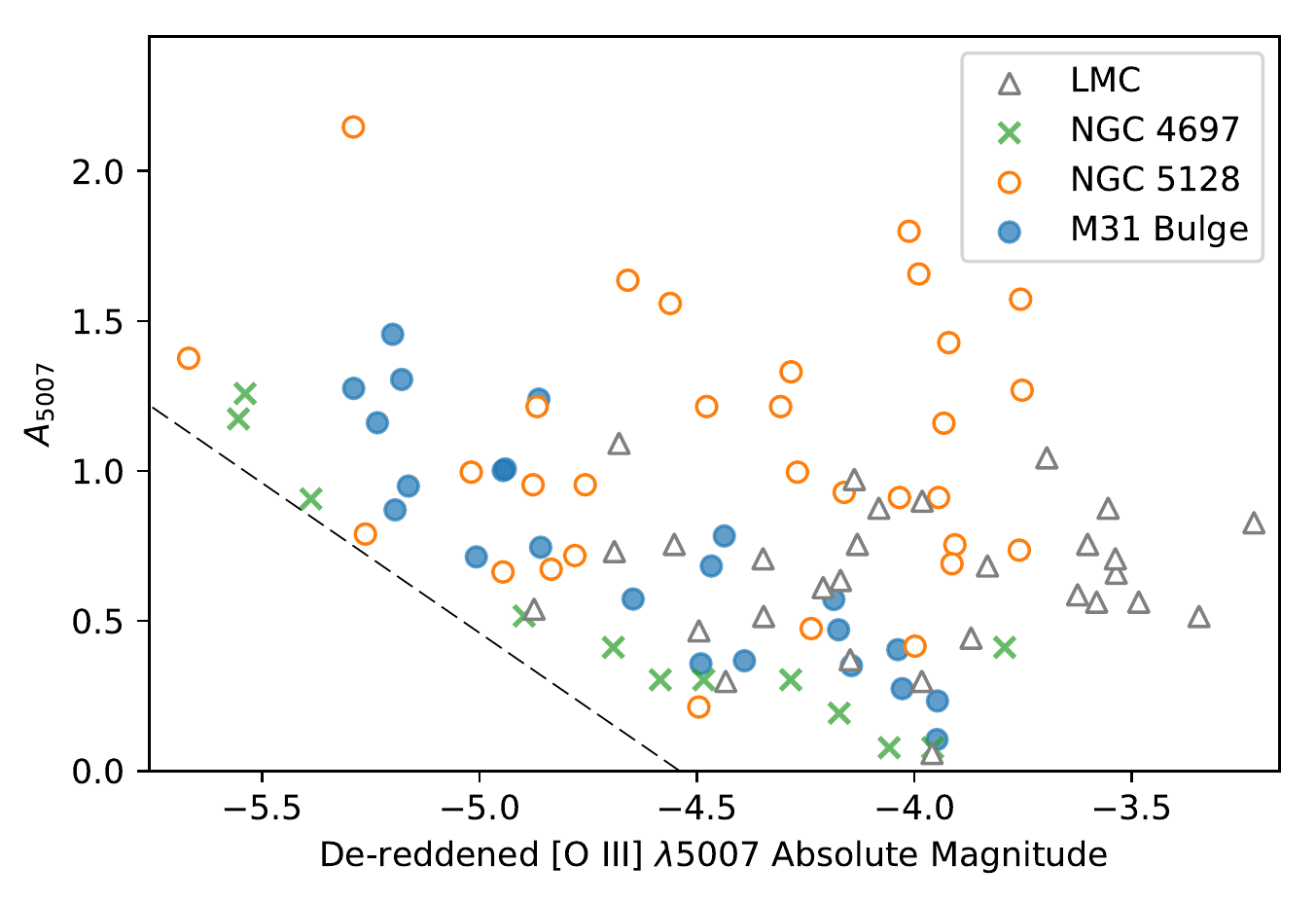}
    \caption{Circumstellar extinctions and de-reddened absolute magnitudes of the brightest PNe in the LMC, NGC~4697, NGC~5128, and the bulge of M31. The dashed line represents the absolute magnitude of the observed PNLF cutoff, $M^* = -4.54$, when accounting for extinction. Selection effects may be responsible for the apparent trends, which are especially strong in NGC~4697 and the bulge of M31. If real, these trends would suggest that intrinsically brighter PNe contain more dust.}
    \label{fig:all_A_vs_M}
\end{figure}

Figure~\ref{fig:all_A_vs_M} compares the circumstellar extinctions and de-reddened absolute magnitudes of the brightest PNe in the LMC \citep[$D \approx 50$ kpc;][]{pejcha2012}, NGC~4697, NGC~5128, and the bulge of M31. Again, each of these samples consists of PNe within the brightest 2 mag of the de-reddened PNLF (which corresponds to the brightest $\sim 1$ mag of the observed PNLF). Since all of the PNe in a given sample come from a relatively narrow range in apparent magnitude, it is natural to expect that the objects with the most circumstellar extinction would have the brightest absolute magnitudes. Hence, it is possible that selection effects are responsible for the observed trends. However, independent of selection effects, the range of de-reddened absolute magnitudes implies that conventional single-star evolution would require a wide range of ages for the progenitors of these objects (as shown in Figure~\ref{fig:all_tMS_hist}) that supposedly derive from older, nearly singular-age populations. This is fundamental to the elliptical galaxy PNLF conundrum, and is the key issue that must be addressed by any model which attempts to explain the PNLF.

\begin{figure}
    \centering
    \includegraphics[scale=0.63]{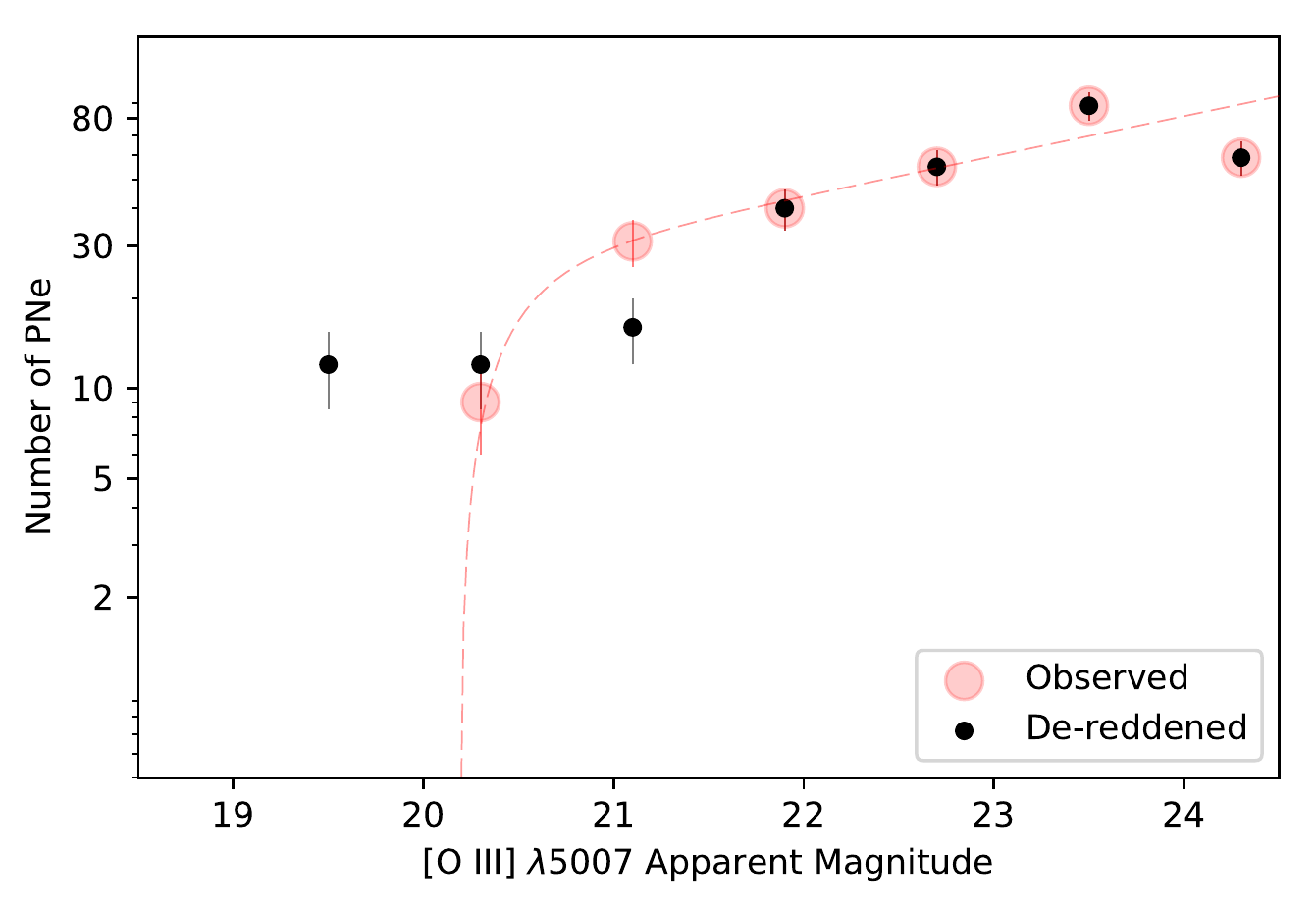}
    \caption{Observed and de-reddened PNLFs for the bulge of M31. The dim ends of the two PNLFs ($m>21$) are identical, as only 23 bright PNe in the sample are de-reddened. The dashed line shows the best fit for the observed data, using Equation~(\ref{eq:N(m)}) with $m^*=20.19$.  While the observed PNLF has a sharp exponential bright-end cutoff, the de-reddened bright end shows no such feature.}
    \label{fig:dered_pnlf}
\end{figure}

Figure~\ref{fig:dered_pnlf} displays the de-reddened bright end of the PNLF in M31's bulge.  Interestingly, the shape of the function shows no evidence of an exponential cutoff, but instead appears to be consistent with a simple power-law distribution. This is possibly due to the fact that only the brightest PNe in the PNLF are de-reddened; if the entire PNLF were to be de-reddened, the bright end might take on a different shape. If there exists some well-behaved distribution of extinctions, as suggested by \cite{reid2010}, a convolution of this distribution with that of a model de-reddened PNLF should reproduce the observed PNLF that has been observed in many galaxies.  We reserve this modeling for a future study. 

Nevertheless, it is interesting that the populations considered in this study have large numbers of PNe with luminosities significantly brighter than $M^*$.  The utility of the PNLF as an extragalactic standard candle rests on the fact that $M^*$ is relatively constant across a very wide range of stellar populations.  Figure~\ref{fig:all_A_vs_M} suggests that this may be due to the systematics of circumstellar extinction, rather than on any property of the PN central stars or the physics of their nebulae.   To test this hypothesis, we would need more reddening measurements for a larger set of statistically complete PN samples.

\section{Discussion}

One possible mechanism for creating high-mass PN central stars in early-type systems is binary star evolution. \cite{ciardullo2005} have argued that conservative mass transfer on the main sequence could produce a blue straggler population with the properties necessary to explain $M^*$ PNe.  In their scenario, pairs of $\sim 1 \, \rm{M}_{\odot}$ stars could be driven to coalescence via the \cite{kozai1962} mechanism or magnetic braking, and thereby produce a continuous stream of $\sim 2 \, \rm{M}_{\odot}$ stars, which could then evolve into $\sim 0.6 \, \rm{M}_{\odot}$ PN central stars.  The lower requisite masses inferred from the \cite{bertolami2016} models assist in this regard.  However, our new results show that in M31's bulge, $\sim 0.66 \, \rm{M}_{\odot}$ central stars are needed to explain $M^*$ PNe, and these objects are generally considered to require $\sim 2.5\,\rm{M}_{\odot}$ progenitors \citep[e.g.,][]{kalirai2008}. 

Can binary evolution produce the extremely luminous PNe seen in old stellar populations? Based on their luminosity-specific number density and estimated lifetimes, \cite{ciardullo2005} suggested that blue straggler stars might be responsible for producing \oiii-bright PNe seen in elliptical galaxies and M31's bulge.  However, the de-reddened luminosities of the M31 bulge PNe present a problem for this scenario: even if two equal-mass $\sim 1.2$~M$_{\odot}$ main sequence stars were to merge (with no mass lost in the interaction), the product would likely still have too low a mass to produce the requisite $\sim 0.66$~M$_{\odot}$ core. Given that the lifetime of a 1.2 M$_{\odot}$ star ($\sim 4$~Gyr) is already straining models for elliptical galaxy stellar populations, it would seem that this simplest of merger scenarios may not be the solution.

Alternatively, one could try to produce super-$M^*$ PNe via a binary interaction on the red giant branch (RGB) or AGB. But even here there are problems. Since low-mass single stars cannot build helium cores more massive than $\sim 0.48$~M$_{\odot}$, \citep{buzzoni1983}, any theory which seeks to create a high-mass PN central star via mass transfer on the RGB must be one which produces a merged remnant, rather than a close binary system. Mass transfer on the AGB has the same problem: it is much easier to build a high-mass PN central star if the common envelope process leads to the creation of a single core, rather than two separate stars. Unfortunately, until our understanding of this phase of binary stellar evolution improves \citep[e.g.,][]{taam2010, ivanova2013},  the question of which of these scenarios can produce super-$M^*$ PNe will remain open.

Given the extremely high masses of our M31 bulge objects, it is reasonable to ask whether these bright \oiii\ sources are planetary nebulae at all. Clearly, their excitation is much larger than that expected from \hii\ regions \citep{shaver1983,kniazev2005,pena2007} or supernova remnants \citep{davis2018}, but direct evidence for their evolutionary status is lacking.  \cite{soker2006} has offered the theory that the brightest \oiii\ sources in early-type galaxies are actually symbiotic stars powered by the accretion of matter onto white dwarfs.  While difficult to rule out, such a model does not answer the question of why the properties of these brightest \oiii\ sources are so similar to those of normal PNe, nor address the question of the relatively high number of objects seen per unit galaxy luminosity.

We note that the luminosity of $M^*$ PNe in old stellar populations is a puzzle that affects more than just the field of post-AGB evolution. Luminous ($M^*$) PNe are rare, in part because they spend so little time ($t < 1$~kyr)  in this phase. However, immediately prior to their evolution blueward in the HR diagram, these PNe were thermally pulsing stars at the tip of the asymptotic giant branch.  These TP-AGB stars can dominate the infrared light of a stellar population, and an understanding of their numbers and evolution is critical for topics such as galaxy population synthesis \citep[e.g.,][]{maraston2006, conroy2009}, and the calibration of the Surface Brightness Fluctuation method \citep[i.e.,][]{liu2000, blakeslee2001, conroy2010}. If stellar evolution models are failing to predict the existence of tens of extremely luminous PNe, then, from the relative lifetimes of the objects, they must also be missing the contribution of thousands of extremely luminous TP-AGB stars \citep{girardi2007, gesicki2018}.  Such an error can lead to an incorrect estimate of a galaxy's stellar mass, star formation rate history, chemical history, and even its initial mass function
\citep[e.g.,][]{conroy2012, conroy2014}. Thus, the issue is one of the more important questions in astronomy.

\section{Conclusions}

In this study, we have investigated the effects of circumstellar extinction on PNe in order to derive the true, de-reddened PNLF of an old stellar population. Our main results and conclusions can be summarized as follows:

\begin{enumerate}
   \item The de-reddened PNLF of M31's inner bulge ($r<3$\arcmin) features PNe that are up to $\sim 1$~mag brighter than the observed PNLF cutoff magnitude, $M^*$, indicating central stars with luminosities $\gtrsim 11{,}000$~L$_{\odot}$ and masses $\gtrsim 0.66$~M$_{\odot}$.
   \item If the initial-final mass relation \citep[e.g.,][]{kalirai2008} holds, such objects must be formed from progenitors with masses $\gtrsim 2.5$~M$_{\odot}$ and evolutionary timescales of less than 1~Gyr. While M31's inner bulge may contain some intermediate-mass stars, this population is not nearly substantial enough to explain the large number of $M^*$ PNe present in the region. 
   \item The de-reddened PN luminosities observed in the bulge of M31 are consistent with those measured in the early-type galaxies NGC~4697 (Hubble type E6) and NGC~5128 (interacting S0p). In these galaxies, central star masses can exceed $0.72$~M$_{\odot}$, implying progenitor masses $\gtrsim 3$~M$_{\odot}$ with main sequence lifetimes $\lesssim 700$~Myr.
   \item Evolutionary models for post-AGB single stars cannot explain the high luminosities of the brightest PNe found in old stellar populations, such as that found in M31's bulge. This discrepancy implies that single-star population synthesis models must be underestimating the maximum luminosities and total integrated starlight associated with the AGB stars of old stellar systems.  
   \item Binary evolution may offers a solution for the problem of super-$M^*$ planetaries, but the simplest scenario, that of mass transfer on the main sequence and the creation of blue straggler stars, is problematic. Common envelope evolution may work, but only if the interaction leads to the creation of a single star, rather than a close binary system.
   \item An explanation remains elusive for the uniformity across stellar populations of $M^*$, the observed bright-end cutoff magnitude of the PNLF.
\end{enumerate}

\acknowledgments
We would like to thank Mallory Molina for her assistance in developing a method for removing M31's diffuse emission from our PNe. We would also like to thank the referee for helping to improve the presentation and arguments made in this paper. This study was supported via NSF through grant AST-1615526\null.  The Institute for Gravitation and the Cosmos is supported by the Eberly College of Science and the Office of the Senior Vice President for Research at the Pennsylvania State University.

\facility{Hobby-Eberly Telescope (LRS2-B)}

\bigskip 

\appendix
\section{QLP}

The LRS2 Quick-Look Pipeline (\texttt{QLP}; Indahl, B.L., in preparation) is an IFU data reduction package specifically designed for the LRS2 spectrographs of the Hobby-Eberly Telescope. \texttt{QLP} finds the correct set of calibration data for each HET observation and uses these frames to produce 1D sky-subtracted spectra. Each spectrograph arm and night is reduced independently, since the calibration data for each is different.  At its core, \texttt{QLP} is a Python wrapper which finds and organizes the data through the various steps of the reduction, while calling other routines to perform the reduction tasks.

At their most basic level, LRS2 data reductions are accomplished with \texttt{Cure}, a software package written in C++, and developed for the Visible Integral-field Replicable Unit Spectrograph \citep[VIRUS;][]{hill2016} of the Hobby-Eberly Telescope Dark Energy Experiment \citep[HETDEX;][]{hill2008}. \texttt{Cure} contains individual tools to perform reduction steps, but is not in itself a reduction pipeline. Instead, \texttt{QLP} calls a specific build of \texttt{Cure}, in which the the CCD size, wavelength ranges, fiber size, and fiber separation are defined.

Each night a set of calibration data is taken at the HET for both LRS2 red and blue units. A facility calibration unit (FCU) feeds light to each spectrograph through one of two liquid light guides, optimized for either the red or the blue. This FCU mimics the focal plane of the telescope and feeds light for fiber flats (using a laser driven light source) and wavelength calibration (HgCd and FeAr arcs). A set of bias frames is also taken each night for calibration.

The \texttt{QLP} wrapper performs its basic reduction by first trimming and subtracting the overscan from each science and calibration frame. Cosmic ray rejection is performed via the \texttt{L.A.Cosmic} algorithm \citep{van2001}, and a master bias is created through the \texttt{Cure} routine, \texttt{meanfits}, which implements kappa sigma clipping.
This master bias is subtracted from the flats, arcs, and science frames. 
	
Each of the 4 CCDs in LRS2 have two amplifiers. First, the data from the two amplifiers are combined by multiplying each amp by its respective gain as recorded in the header. The arcs and flats are then normalized by their exposure times and combined through \texttt{meanfits}.  The master flat and master arc images are then used to define the fiber traces and build the wavelength solutions.  Flat fielding is not performed at this stage, but pixel-to-pixel variations are accounted for through weights which were derived while building the distortion solution. These weights are implemented in the sky subtraction and fiber extraction routines.
  
After the data from the 4 CCDs are combined, the \texttt{Cure} routine \texttt{Deformer} is called to locate the positions of the fibers in the images, build their profile perpendicular to the dispersion direction, and create their wavelength solutions. As with any IFU dataset, each fiber has a different curvature and profile in the image. \texttt{Deformer} uses the master trace, the master arc, and a list of arc lines to compute the traces and wavelength solution for each fiber via a series of seventh-order Chebychev polynomials.    The curves translate $(x,y)$ positions on the combined CCD to $(w,f)$ coordinates, where $w$ represents the direction of constant wavelength, and $f$ moves in the direction of the fiber.

\texttt{Deformer} also uses an initial guess for a Gauss-Hermite expansion to fit overlapping profiles to three fibers at a time.  Each solution is used as an initial guess for the next set of fibers, and the process is iterated until it converges on a solution. Chebychev polynomials are then fit over the variation in the Gauss-Hermite fits. This fiber model provides statistical weights which describe how much flux from each fiber contributes to each pixel. The weights described by this model essentially provide the flat field as they are used in the fibers extraction and in later \texttt{Cure} routines.

Sky subtraction is accomplished using the \texttt{Cure} routine \texttt{subtractsky}. In its recommended mode, \texttt{subtractsky} uses fibers in the science frames which are devoid of continuum to model the sky emission. A spectral window 200 pixels wide is selected, and within that window, each pixel's flux is divided by its weight from the fiber model, and by the Jacobian of the transform, into wavelength space to account for the changing dispersion along a fiber. Pixels are then sorted by their wavelengths, and at each approximate wavelength, outliers are filtered out. A B-spline is fit along the wavelength direction, and spectra with a S/N minimum of 15 above the spline are flagged to contain continuum and are excluded in the following iteration.  The resulting B-spline value at the corresponding wavelength for each pixel is considered to represent the sky and is then multiplied by that pixel's weight and Jacobian. 

The final step of the basic reduction uses two \texttt{Cure} routines: \texttt{fiberextract} and \texttt{mkcube}. The \texttt{fiberextract} routine uses the fiber model to fit the fluxes of each fiber, so that flux from one fiber does not contaminate that from the adjacent fibers. The \texttt{mkcube} routine then builds the spectra from the 280-fiber array into a data cube with dimensions of $42 \times 24$ pixels on the sky (by interpolating over this spatial grid with a Gaussian kernel), and 2,000 pixels in the wavelength direction.

\end{document}